\let\oldvec\vec
\documentclass{article}
\usepackage[affil-it]{authblk}
\let\vec\oldvec
\usepackage[OT1]{fontenc}

\usepackage[a4paper, bottom=2.75cm,left=2.25cm, right=2.25cm,top=2.75cm]{geometry}
\usepackage{graphicx}
\usepackage{amssymb,amsmath}
\usepackage{txfonts}
\usepackage[numbers]{natbib}
\usepackage{hyperref}
\usepackage{rotating}
\usepackage{color, colortbl}
\usepackage{physics}
\usepackage[section]{placeins}

\usepackage{bibunits}

\usepackage{etoolbox}

\makeatletter
\newcommand*{\newbibstartnumber}[1]{%
  \apptocmd{\thebibliography}{%
    \global\c@NAT@ctr #1\relax
    \addtocounter{NAT@ctr}{-1}%
  }{}{}%
}
\makeatother

\usepackage{filecontents}

\renewcommand{\vec}[1]{\mathbf{#1}}

\definecolor{LightCyan}{rgb}{0.88,1,1}

\definecolor{Gray}{gray}{0.9}

\DeclareSymbolFont{mymathvariables}{OT1}{cmr}{m}{n}
\SetSymbolFont{mymathvariables}{normal}{OT1}{cmr}{m}{n}
\DeclareSymbolFontAlphabet{\mathnormal}{mymathvariables}
\DeclareMathSymbol{v}{\mathalpha}{mymathvariables}{118}

\newlength{\wfig}
\newlength{\wfi}
\newlength{\wfigure}
\makeatletter
\let\jnl@style=\rm
\def\ref@jnl#1{{\jnl@style#1}}

\def\aj{\ref@jnl{AJ}}                   
\def\actaa{\ref@jnl{Acta Astron.}}      
\def\araa{\ref@jnl{ARA\&A}}             
\def\apj{\ref@jnl{ApJ}}                 
\def\apjl{\ref@jnl{ApJ}}                
\def\apjs{\ref@jnl{ApJS}}               
\def\ao{\ref@jnl{Appl.~Opt.}}           
\def\apss{\ref@jnl{Ap\&SS}}             
\def\aap{\ref@jnl{A\&A}}                
\def\aapr{\ref@jnl{A\&A~Rev.}}          
\def\aaps{\ref@jnl{A\&AS}}              
\def\azh{\ref@jnl{AZh}}                 
\def\baas{\ref@jnl{BAAS}}               
\def\bac{\ref@jnl{Bull. astr. Inst. Czechosl.}}
\def\caa{\ref@jnl{Chinese Astron. Astrophys.}}
\def\cjaa{\ref@jnl{Chinese J. Astron. Astrophys.}}
\def\icarus{\ref@jnl{Icarus}}           
\def\jcap{\ref@jnl{J. Cosmology Astropart. Phys.}}
\def\jrasc{\ref@jnl{JRASC}}             
\def\memras{\ref@jnl{MmRAS}}            
\def\mnras{\ref@jnl{MNRAS}}             
\def\na{\ref@jnl{New A}}                
\def\nar{\ref@jnl{New A Rev.}}          
\def\pra{\ref@jnl{Phys.~Rev.~A}}        
\def\prb{\ref@jnl{Phys.~Rev.~B}}        
\def\prc{\ref@jnl{Phys.~Rev.~C}}        
\def\prd{\ref@jnl{Phys.~Rev.~D}}        
\def\pre{\ref@jnl{Phys.~Rev.~E}}        
\def\prl{\ref@jnl{Phys.~Rev.~Lett.}}    
\def\pasa{\ref@jnl{PASA}}               
\def\pasp{\ref@jnl{PASP}}               
\def\pasj{\ref@jnl{PASJ}}               
\def\rmxaa{\ref@jnl{Rev. Mexicana Astron. Astrofis.}}%
\def\qjras{\ref@jnl{QJRAS}}             
\def\skytel{\ref@jnl{S\&T}}             
\def\solphys{\ref@jnl{Sol.~Phys.}}      
\def\sovast{\ref@jnl{Soviet~Ast.}}      
\def\ssr{\ref@jnl{Space~Sci.~Rev.}}     
\def\zap{\ref@jnl{ZAp}}                 
\def\nat{\ref@jnl{Nature}}              
\def\iaucirc{\ref@jnl{IAU~Circ.}}       
\def\aplett{\ref@jnl{Astrophys.~Lett.}} 
\def\apspr{\ref@jnl{Astrophys.~Space~Phys.~Res.}}
\def\bain{\ref@jnl{Bull.~Astron.~Inst.~Netherlands}}
\def\fcp{\ref@jnl{Fund.~Cosmic~Phys.}}  
\def\gca{\ref@jnl{Geochim.~Cosmochim.~Acta}}   
\def\grl{\ref@jnl{Geophys.~Res.~Lett.}} 
\def\jcp{\ref@jnl{J.~Chem.~Phys.}}      
\def\jgr{\ref@jnl{J.~Geophys.~Res.}}    
\def\jqsrt{\ref@jnl{J.~Quant.~Spec.~Radiat.~Transf.}}
\def\memsai{\ref@jnl{Mem.~Soc.~Astron.~Italiana}}
\def\nphysa{\ref@jnl{Nucl.~Phys.~A}}   
\def\physrep{\ref@jnl{Phys.~Rep.}}   
\def\physscr{\ref@jnl{Phys.~Scr}}   
\def\planss{\ref@jnl{Planet.~Space~Sci.}}   
\def\procspie{\ref@jnl{Proc.~SPIE}}   

\makeatother

\begin{document}

\title{Fine structure of type III solar radio bursts from Langmuir wave motion in turbulent plasma}

   \author{Hamish A. S. Reid$^{1,2}$ and Eduard P. Kontar$^1$}
    \affil{\normalsize $^1$School of Physics and Astronomy,   University of Glasgow, G12 8QQ, United Kingdom \\ 
     $^2$Department of Space and Climate Physics, University College London, RH5 6NT, United Kingdom}



\maketitle
The Sun frequently accelerates near-relativistic electron beams that travel out through the solar corona and interplanetary space. Interacting with their plasma environment, these beams produce type III radio bursts, the brightest astrophysical radio sources seen from the Earth. The formation and motion of type III fine frequency structures is a puzzle but is commonly believed to be related to plasma turbulence in the solar corona and solar wind. Combining a theoretical framework with kinetic simulations and high-resolution radio type III observations using the Low Frequency Array, we quantitatively show that the fine structures are caused by the moving intense clumps of Langmuir waves in a turbulent medium. Our results show how type III fine structure can be used to remotely analyse the intensity and spectrum of compressive density fluctuations, and can infer ambient temperatures in astrophysical plasma, both significantly expanding the current diagnostic potential of solar radio emission.

\vspace{\baselineskip}


\section{Introduction}

The Sun routinely accelerates electrons in its outer atmosphere
that subsequently travel through interplanetary space.
Electron beams generate Langmuir waves in the background plasma,
producing bright solar radio emission \citep{Suzuki:1985aa,Pick:2008aa,Reid:2014aa}.
Thanks to our close proximity to the Sun, type III radio bursts are the brightest radio sources in the sky, providing a unique opportunity to understand particle acceleration and transport \citep{Aschwanden:2005aa,Holman:2011aa}. Type IIIs also provide us remote diagnostics of solar corona and solar wind properties, with stellar radio bursts having the diagnostic potential for stellar atmospheres \citep{Dulk:1985aa}.
Type III bursts present a remote indicator for energetic electrons
escaping the Sun which is important for our prediction
of extreme space weather events \citep{Posner:2007aa}.

As electron beams propagate through plasma with decreasing density, and hence plasma frequency,
type III bursts drift from high to low frequencies in dynamic spectra (frequency vs time plots).  Type III frequency drift is often used to infer the bulk velocity of electron beams which drive type III emission.  Fine frequency structures are often present in dynamic spectra of type III bursts, observed as nearly horizontal stria in the envelope of a type III burst \citep{de-La-Noe:1972aa,Abranin:1978aa,Kontar:2017aa,MelNik:2018aa}.
The stria have short duration (e.g. ~1 sec at 30 MHz)  and have characteristic frequency fraction width $\Delta f/f \approx 0.1 $, independent of frequency.
However, the properties of these striae are puzzling
and not understood with a large number of competing theories \citep{de-La-Noe:1972aa,Melrose:1983aa,Kolotkov:2018aa}.

The drift rate of individual stria give derived speeds around $0.6~\rm{Mm~s}^{-1}$  \citep{Sharykin:2018aa}.
Intriguingly, these velocities are significantly smaller than electron beam velocities
of $100~\rm{Mm~s}^{-1}$ \citep{Suzuki:1985aa},
but much greater than coronal sound speeds
of $\sqrt{kT/M}\simeq 0.2~\rm{Mm~s}^{-1}$, where $M$
is the ion mass, $kT$ is the plasma temperature in energy units \citep{Pecseli:2012aa}.
It is therefore not clear what plasma process dictates striae drift rates
and the origin and quantitative explanation of type III burst striae
presents a major challenge.  The power spectrum of type III stria have been shown for one event to have a roughly 5/3 power-law slope both for fundamental and harmonic emission \citep{Chen:2018aa}.
It is likely that the striae are related to some compressive density perturbations
\citep{de-La-Noe:1972aa}, hence can provide important insights
into the magnetohydrodynamic (MHD) waves at heights where extreme ultraviolet (EUV) observations are normally not available.  
To establish the progenitor of striae in dynamic spectrum,
one needs high frequency resolution observations combined
with the numerical simulations of the solar radio bursts.

\begin{figure}
\center
\includegraphics[width=\wfigure]{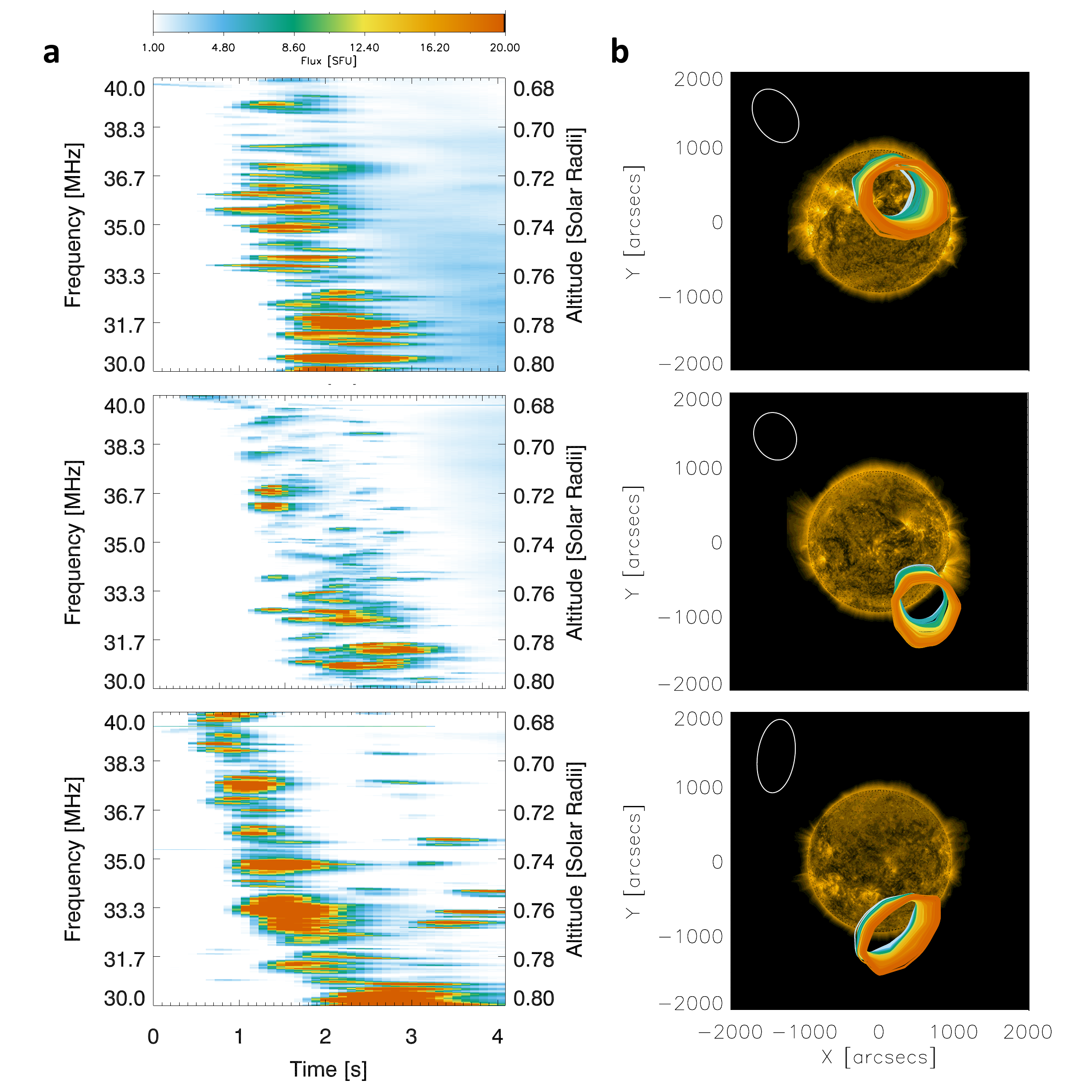}
\caption{\textbf{Dynamic spectra and associated radio contours of solar type III radio bursts.}  \textbf{a:} dynamic spectra of the flux [s.f.u.] for fundamental type III bursts showing fine structure.  Events from top to bottom occurred on the 16th April 2015 at 11:56:20 UT, 24th June 2015 at 12:18:20 UT and 16th September 2015 at 09:56:33 UT.  The corresponding solar altitude assuming the Parker density model \citep{Parker:1958aa} is also shown parallel to the frequency.  \textbf{b:}  The LOFAR contours at $75\%$ of the peak flux of the type III bursts going from 40 MHz to 30 MHz in the colour sequence white-blue-green-yellow-red.  The LOFAR beam contour at $75\%$ for 30 MHz is shown in the top left corner in white.  The background is the Sun in EUV at 171 Angstroms observed by AIA.}
\label{figure:typeIII_dynspec}
\end{figure}

In this study, we demonstrate that the observed properties
of type III bursts fine structure are consistent with moving Langmuir waves
through turbulent space plasma.
Furthermore, the drift rate of the stria obtained in the observations
is consistent with the group velocity of the Langmuir waves.
The numerical simulation can reproduce high resolution LOFAR observations
of type III fine structure in the solar corona. This more complete understanding
of striae opens a new opportunity for diagnostics of density fluctuations parallel to the magnetic field within the solar corona, which are not available otherwise.

\section{Type III observations} \label{t3:obs}

We present three examples of type III bursts showing obvious fine structure, observed using the LOw Frequency ARray (LOFAR) \citep{van-Haarlem:2013aa} on the 16th April, 24th June and 16th September 2015.
Figure \ref{figure:typeIII_dynspec} shows calibrated dynamic spectra of the fundamental emission from the different events observed between 30--40~MHz.  Analysis of observational properties from the event on the 16th April 2015 has been previously reported by \citep{Kontar:2017aa,Chen:2018aa,Sharykin:2018aa,Kolotkov:2018aa}.
Each event lasts a few seconds at these frequencies and
all events had fundamental-harmonic pairs similar to the 16th April event
that was shown by \citep{Kontar:2017aa}.
The large-scale frequency drift from high to low frequencies
provides an estimate of electron beam bulk velocity (see Methods) of $v_b=88~\rm{Mm~s}^{-1}$,  $49~\rm{Mm~s}^{-1}$, $46~\rm{Mm~s}^{-1}$, for the three events, in chronological order.
Figure \ref{figure:typeIII_dynspec} also shows the LOFAR radio contours
at different radio frequencies plotted over the extreme ultraviolet (EUV) 171~\AA~Sun, observed by the Atmospheric Imaging Assembly (AIA) \citep{Lemen:2012aa}.  The radio contour for each frequency channel is displayed at the times corresponding to the peak flux.  The radio contours show how a solar-accelerated electron beam produces lower frequency radio emission as it propagates outwards from the Sun.

Fine structure in the fundamental emission is evident in the dynamic spectra (Fig. \ref{figure:typeIII_dynspec}), with clumps of radio emission appearing at varying intensities. We can quantify the characteristic intensity of the fine structure, the relative flux fluctuation amplitude $\Delta I/I$, between the frequencies 30--40~MHz for all three events (see Methods).  Taking the average variation from the mean intensity (Figure \ref{figure:typeIII_peakflux}), we found typical values of $\Delta I/I=1.41, 1.01, 1.35$  for the three events, in chronological order.  High values for $\Delta I/I$ are expected given that we selected these events for their abundant fine structure.  However, if we want to use the intensity of radio fine structure as a plasma diagnostic tool, we must understand their origins.

\section{Radio fine structure driver}

The cause of type III fine structure has previously been thought to relate to modulations in the growth rate of beam-induced Langmuir waves, naturally leading to a modulation in the intensity of radio waves.  Langmuir wave growth rates can be modulated via wave refraction, which is controlled by the level of density fluctuations in the background plasma \citep[e.g.][]{Reid:2010aa,Li:2012aa,Loi:2014aa,Reid:2017ab}. Previous simulations \citep{Li:2012aa,Loi:2014aa} were able to produce synthetic type III dynamic spectra with fine structure at frequencies corresponding to density variations in the background plasma.  Some of the synthetic dynamic spectra showed discrete structures in frequency.  However, the studies did not reproduce any frequency drift associated with these structures and the number of structures, particularly in fundamental emission, was much smaller than the number observed in type III striae bursts.  The idea used by the preceding studies and other related works, that density fluctuations can modulate Langmuir waves and causes radio fine structure, has been discussed for decades \citep[e.g.][as a review]{Melrose:1986aa}.  However, nobody has created a robust theoretical model that allows one to reverse engineer the level and spectrum of density fluctuations from the observed radio fine structure.  Such a feat enables type III radio bursts to be used as a probe of space plasma density turbulence where in situ observations are not feasible.

We have shown (see Methods) that the motion of Langmuir waves
travelling with a group velocity $v_{\rm gr}=3v_{\rm Th}^2/v_b$
has a shift $\delta v$ in Langmuir wave phase velocity from refraction off density fluctuations with intensity $\Delta n/n$.  The shift $\delta v$ causes the Langmuir waves to move out of resonance with the electron beam, suppressing wave growth.  The intensity $\Delta n/n$ can directly be related to the intensity of radio fine structure $\Delta I/I$ via
\begin{equation}\label{eqn:didn}
\frac{\langle\Delta n^2\rangle}{n^2} = \left(\frac{v_{\rm{Th}}^2}{v_b^2}\right)^2\frac{\langle\Delta I^2\rangle}{I^2}
\end{equation}
where $v_{\rm Th}=\sqrt{k_bT_e/m_e}$ is the background thermal velocity and $v_b$ is the velocity in the electron beam distribution (see Methods).
The above equation has broad implications, not only explaining how radio fluctuations are related to the turbulent intensity of the emitting astrophysical plasma, but also showing that the radio fluctuations are the signatures of Langmuir wave motion.  The velocities that dictate the intensity of radio fluctuations can be viewed as a ratio between electron beam
and Langmuir wave spatial velocities $v_{\rm Th}^2/v_b^2=v_{\rm gr}/(3v)$.  Equation \ref{eqn:didn} quantifies how radio fluctuations can be a powerful remote probe of plasma density fluctuations in the solar (or stellar) atmosphere and wind.

\begin{figure}
\center
\includegraphics[width=\wfigure]{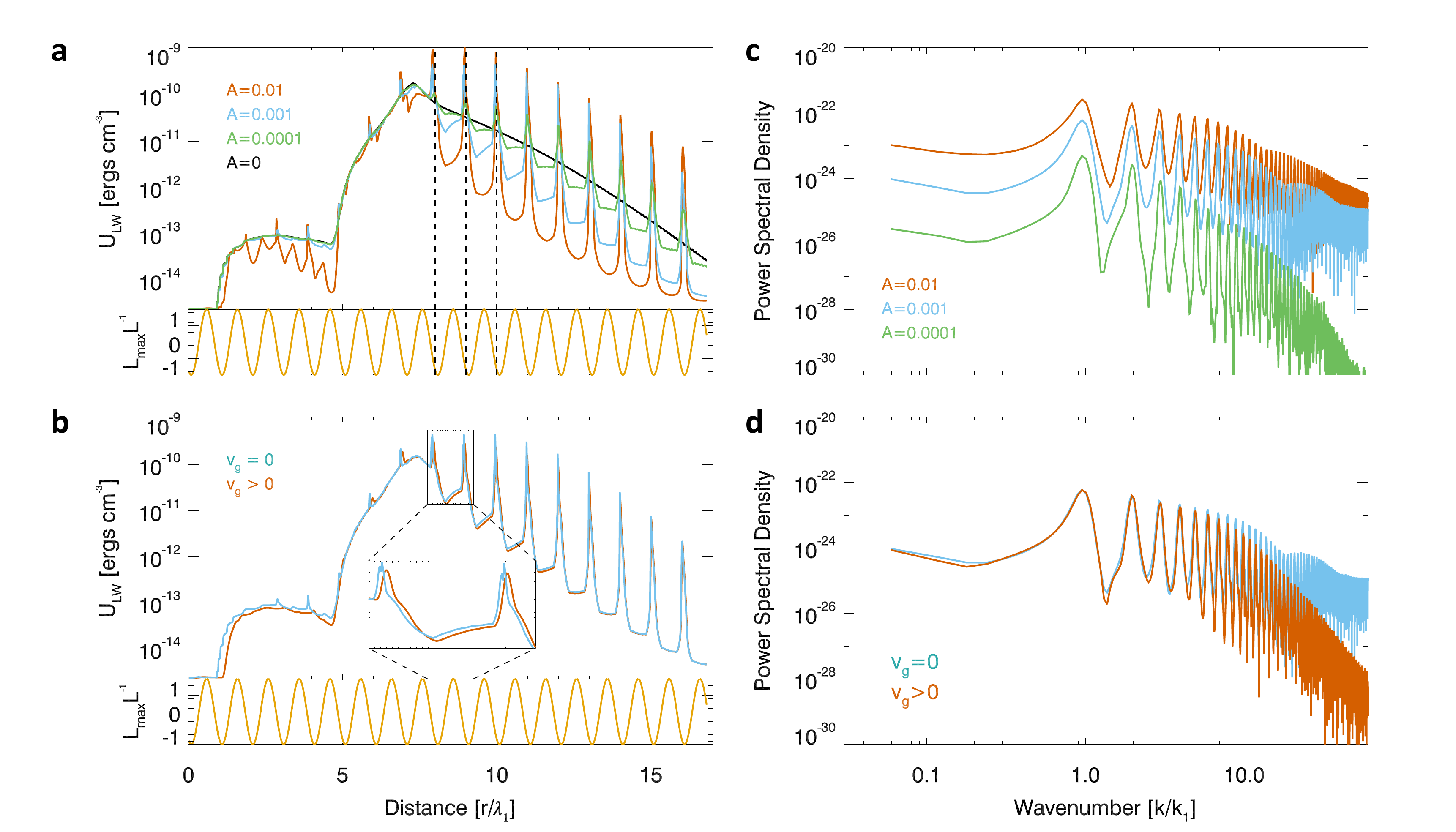}
\caption{\textbf{Simulated beam-generated Langmuir wave energy densities and their associated power density spectra.} \textbf{a:}  Beam-generated Langmuir wave energy density $U_{LW}$ as a function of distance.  The background plasma has a sinusoidal perturbation of wavelength $\lambda_1=50$~Mm with amplitudes $A=10^{-2},10^{-3},10^{-4},0$, represented by red, blue, green and black, respectively.  The lower panel highlights the modulation in the background plasma gradient length scale $L=0.5n_e^{-1}dn_e/dx$, normalised to one.  The vertical black dashed lines highlight that the peaks of Langmuir wave energy occurs regularly when $L$ is near minimum. \textbf{b:} Beam-generated Langmuir wave energy density for $A=10^{-3}$ without (blue) and with (red) Langmuir wave group velocity simulated.  Note the phase displacement in the Langmuir waves caused by the group velocity highlighted by the magnified box.  \textbf{c:} Power density spectrum of the Langmuir wave energy density minus the unperturbed case ($A=0$) from the simulations shown in the top left.  The peaks occur at the wavenumber $k_1=2\pi/\lambda~\rm{Mm}^{-1}$ for $\lambda_1=50$~Mm, and the harmonics.  \textbf{d:} as panel \textbf{c} for the simulations shown in panel \textbf{b}, with and without Langmuir wave group velocity.  Note the power density spectra at larger $k$ (smaller length scales) is reduced when $v_g>0$.}
\label{figure:lwed_fft_sin_200}
\end{figure}

\section{Langmuir wave modulation} \label{sec:lwaves}
We now show how key parameters of density fluctuation amplitude and wavelength affect the level of beam-driven Langmuir waves, and hence the type III radio bursts.  We also show how the spatial motion of Langmuir waves at the group velocity also affects the level of beam-driven Langmuir waves

We demonstrate this behaviour using 1D kinetic simulations of an electron beam that is injected into the solar corona and then propagates out from the Sun.  The simulations take into account the resonant interactions of electrons with Langmuir waves and other relevant physical mechanisms for electron propagation.  The simulations model is described in the Methods section and is similar to previously used models model \citep[e.g.][]{Reid:2017ab}.  The main difference for the simulations used in this section is that we have a constant background density of $n_0 = 1.1\times10^{7}~\rm{cm}^{-3}$ which corresponds to a plasma frequency of 30 MHz.  This background density is then modulated by a single sinusoidal fluctuations $n_e(x) = n_0(1+ A\sin(k_1x + \phi))$, where $A$ is the amplitude of the density fluctuations,
$k_1=2\pi/\lambda_1$ and $\lambda_1=50$~Mm are the wavenumber
and wavelength of the density fluctuation.  We have used this unrealistic background density to remove the large scale radial decrease in coronal background electron density so we can highlight the effects of the amplitude and wavelength of the sinusoidal density fluctuation.  The real solar corona also has turbulent density fluctuations, which we simulate in Section \ref{sec:t3sims}.

The simulations include additional physical terms not considered for the derivation of Equation \ref{eqn:didn}, such as electron transport and the electron velocity diffusion due to Langmuir wave generation.  The simulations highlight two important relations needed for understanding type III radio fine structure; how Langmuir wave growth mirrors the density fluctuations parameters, and how spatial Langmuir wave motion alters Langmuir wave growth.

The relation between density fluctuation parameters and Langmuir wave growth is best shown by setting the Langmuir wave group velocity to zero.  Figure \ref{figure:lwed_fft_sin_200} shows a snapshot of the Langmuir wave energy density $U_{\rm{LW}}$ at $t=7.5$~s for four simulations with density fluctuation amplitudes $A=0,10^{-4},10^{-3},10^{-2}$.  The wavelength in Langmuir wave fluctuations are the same as the wavelength of the gradient of the background density fluctuations $\partial n_e/\partial x$, indicated by the vertical dashed lines in Figure \ref{figure:lwed_fft_sin_200}.  Increasing the amplitude of the density fluctuations increases the modulation of Langmuir waves. Langmuir wave energy density modulation is further highlighted
by the background-subtracted power spectrum density in Figure \ref{figure:lwed_fft_sin_200}.  The peak in each power spectrum occurs at $k_1=2\pi/\lambda_1~\rm{Mm}^{-1}$ for $\lambda_1=50$~Mm, and the amplitude is governed by the amplitude of the sinusoidal fluctuation.  Modifying the wavelength $\lambda$ of the sinusoidal fluctuation has a corresponding effect on the Langmuir wave energy density, shifting the peak in the Langmuir wave power spectral density to  $k=2\pi/\lambda~\rm{Mm}^{-1}$, shown in the Methods section.

When we include the spatial motion of Langmuir waves motion at the group velocity, the connection between the density fluctuations and the Langmuir wave oscillations is modified. Langmuir waves move away from regions in space which have favourable growth conditions at the group velocity.  This has the effect of smoothing the Langmuir wave
energy at scales smaller than their lifetime multiplied by the group velocity.
Figure \ref{figure:lwed_fft_sin_200} illustrates this behaviour
for two simulations with and without group velocity.
The power is reduced at higher wavenumbers where the Langmuir waves had time to disperse.  However, at lower wavenumbers the power remains the same.  Reducing the wavelength to e.g. $\lambda_2=\lambda_1/5$ causes the power
to be reduced for all the peaks in the spectrum,
as the Langmuir waves exist for long enough that their motion smooths
the wave energy beyond one wavelength of the density fluctuation.

\section{Type III simulations} \label{sec:t3sims}

\begin{figure}
\center
\includegraphics[width=\wfig]{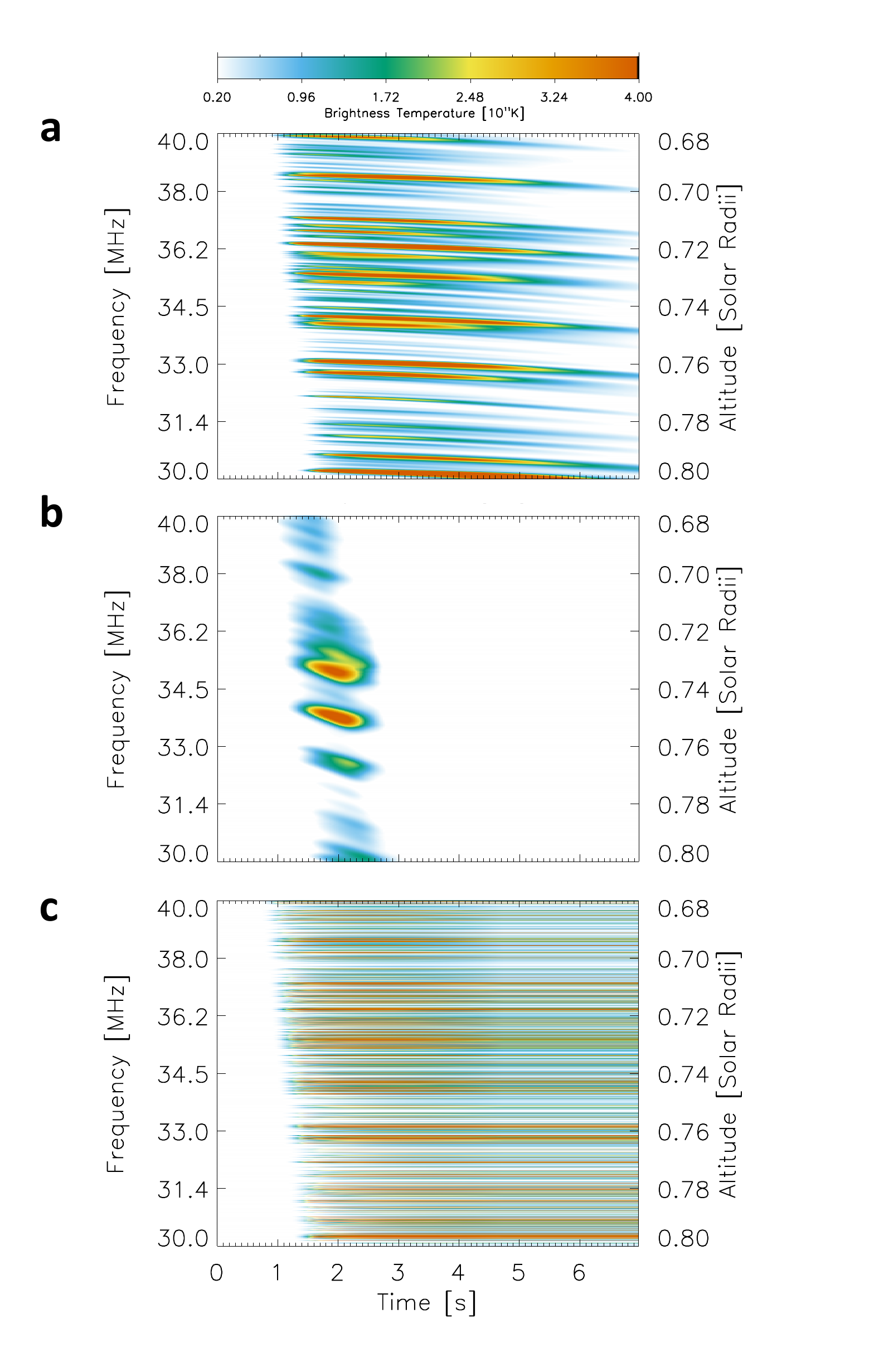}
\caption{\textbf{Simulated type III dynamic spectra.} \textbf{a:} Simulated dynamic spectra showing the fundamental emission between 30 and 40 MHz from a beam travelling through the 1 MK corona.  \textbf{b:} As panel {a} but with a beam travelling through the 10 MK corona.  \textbf{c:} As panel {a} but with no group velocity simulated.}
\label{figure:dynspec_sun}
\end{figure}

Solar-accelerated electron beams propagating out through the solar corona and solar wind can naturally produce type III radio emission with fine structure.  Our kinetic simulations modelled the injection and propagation of an electron beam through the 1 MK solar corona that has a turbulent -5/3 spectrum (Kolmogorov) of coronal density fluctuations (see Methods).  The beam-driven Langmuir waves generate radio emission through the fundamental plasma emission mechanism.  The resulting synthetic type III radio burst dynamic spectrum is shown in Figure \ref{figure:dynspec_sun} between the frequencies 30--40 MHz.  The radio fine structure of the striae is evident.  As demonstrated previously, the Langmuir wave level is affected by Langmuir wave refraction off density fluctuations in the background plasma.  Intense populations of Langmuir waves then move through the solar corona, causing the individual stria.

Properties of the striae are strongly dictated by the temperature of the background plasma.  Figure \ref{figure:dynspec_sun} shows a dynamic spectrum generated by the same electron beam propagating through a hotter 10 MK plasma.  The higher temperature, and hence higher thermal velocity relates to a faster Langmuir wave group velocity, $v_{\rm gr}=3v_{\rm Th}^2/v$, than for a 1 MK plasma.  The radio fine structure is noticeably different.  Faster Langmuir wave motion blurs out striae with small frequency widths, increasing individual stria frequency width and drift rate.  This increased striae frequency width and the striae drift magnitude makes the simulated striae incompatible with the LOFAR observations, consistent with 10 MK being an unrealistically high temperature for the upper solar corona.
The duration of striae is also shortened as higher temperature plasma has increased Landau damping of Langmuir waves generated at the back of the beam.
This shortens the length of time radio is emitted at a given frequency (position).

Langmuir wave motion is absolutely essential to replicate the observed characteristic type III fine structure features.  Figure \ref{figure:dynspec_sun} demonstrates that, when Langmuir wave motion was set to zero ($v_{\rm g}=0$), the dynamic spectrum is dramatically modified.  Wave growth is still modulated by local density gradients but the clumps of intense Langmuir waves do not propagate through space.  Consequently, the fine structure has a much smaller width in frequency space and shows no motion in time, in stark contrast to type III observations.

\begin{figure}
\center
\includegraphics[width=\wfig]{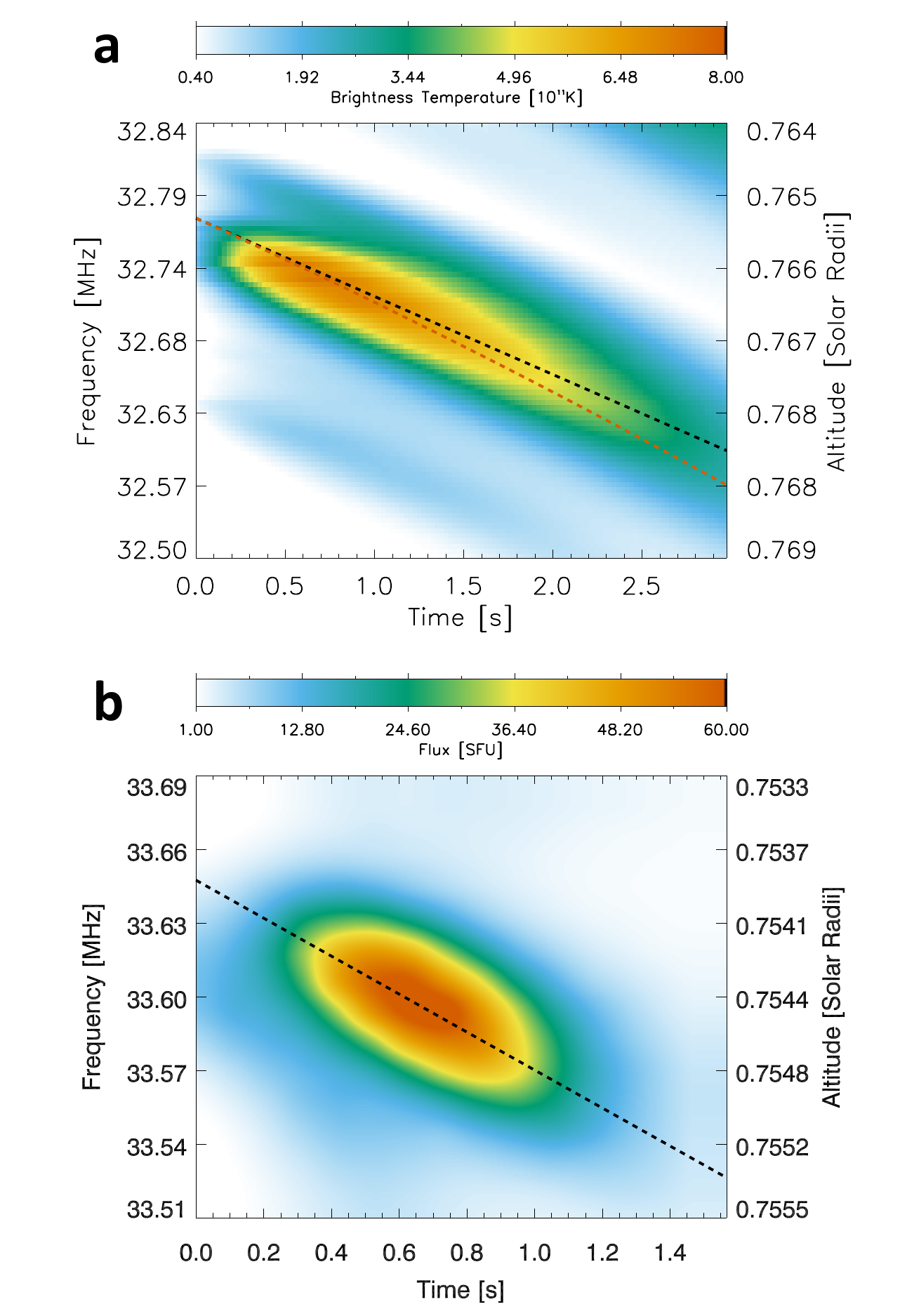}
\caption{\textbf{Magnification of an observed and simulated solar type III stria.} \textbf{a:} A magnification of one stria from the simulated dynamic spectrum where the background temperature was 1 MK.  The black dashed line shows a linear fit, estimating a constant stria velocity of $0.6~\rm{Mm~s}^{-1}$.  The red dashed line estimates a quadratic fit, with an initial stria velocity of $0.63~\rm{Mm~s}^{-1}$, evolving to $0.76~\rm{Mm~s}^{-1}$ after 3 seconds.  \textbf{b:} a magnification of a stria from the 16th April event observed by LOFAR.  The black dashed line shows a linear fit of $0.69~\rm{Mm~s}^{-1}$.  The data has been interpolated to increase the resolution, for clarity.}
\label{figure:striae_vel}
\end{figure}

Clumps of Langmuir waves moving at the group velocity consistently answers the outstanding puzzle of which physical process caused the derived velocity obtained from the frequency drift of individual stria.  Langmuir waves propagate significantly slower than their resonant electrons,
with relevant group velocities around $0.5$ to $4~\rm{Mm~s}^{-1}$ for a 1 MK plasma, compared to electron speeds around $100~\rm{Mm~s}^{-1}$.
Figure \ref{figure:striae_vel} shows the frequency drift
from an individual stria observed by LOFAR.
The linear fit estimates a group velocity of $0.69~\rm{Mm~s}^{-1}$
assuming the Parker density model \citep{Parker:1958aa}.
Figure \ref{figure:striae_vel} also shows an individual stria
from the simulations where the background temperature was 1 MK (top panel of Figure \ref{figure:dynspec_sun}).  The linear fit of $0.69~\rm{Mm~s}^{-1}$ is very similar to the observed stria frequency drift rate, and is identical to the mean group velocity of the Langmuir waves responsible in the simulation.  A quadratic fit indicates a change in the Langmuir wave group velocity with time; relating to the resonating electron beam velocities decreasing from the front to the back of the beam \citep{Reid:2018ab}.  We cannot resolve any significant evolution in the observed stria frequency drift rate.  However, the time duration for the observed stria is about 1 second smaller than the simulated stria.  The stria time duration is likely related to the spread in electron velocities between the fastest and slowest electrons that excite the Langmuir waves.  An increase in the Landau damping from a background plasma with a kappa-like distribution is stronger \citep{Li:2014aa}, is more realistic of the solar wind, and likely to reduce stria duration.  A decrease in the electron beam velocity spectral index would also cause a smaller spread in electron velocities and could reduce stria duration, in a similar way to how it reduces type III duration \citep{Li:2013aa,Reid:2018ab}.

Knowing that the frequency drift of striae is driven by the Langmuir wave group velocity creates a powerful diagnostic tool for plasma temperature. Striae frequency drift contains information about the gradient of the background electron density, the velocity of the electron beam and the background thermal velocity such that
\begin{equation}
\pdv{f_s}{t} = \frac{f}{2n_e}\pdv{n_e}{x}\frac{3v_{\rm{Th}}^2}{v_b},
\end{equation}
where $\partial f_s / \partial t$ is the striae drift rate and $\partial n_e / \partial x$ is the local density gradient. Velocity estimations \citep{Sharykin:2018aa} using striae from the type III on the 16th April 2015 at 11:56 , assuming the Newkirk coronal density model \citep{Newkirk:1961aa}, give an inferred velocity of $v_{\rm gr}=0.58~\rm{Mm~s}^{-1}$ around 40 MHz.  With the beam velocity estimated in Section \ref{t3:obs} as $v_b=88~\rm{Mm~s}^{-1}$ using the Parker density model, we can obtain an estimate of the thermal velocity  $v_{\rm{Th}}=\sqrt{v_{\rm gr}v_b/3}=4.1~\rm{Mm~s}^{-1}$, corresponding to a background temperature of 1.1~MK.  There is a small element of uncertainty in the derived temperature related to the background density model used.  However, there are observational bounds for acceptable density models which limit these uncertainties.  If instead we estimate the beam velocity using the Newkirk density model, we obtain a value of $v=118~\rm{Mm~s}^{-1}$ which corresponds to a background temperature of 1.5 MK.

\begin{figure}
\center
\includegraphics[width=\wfigure]{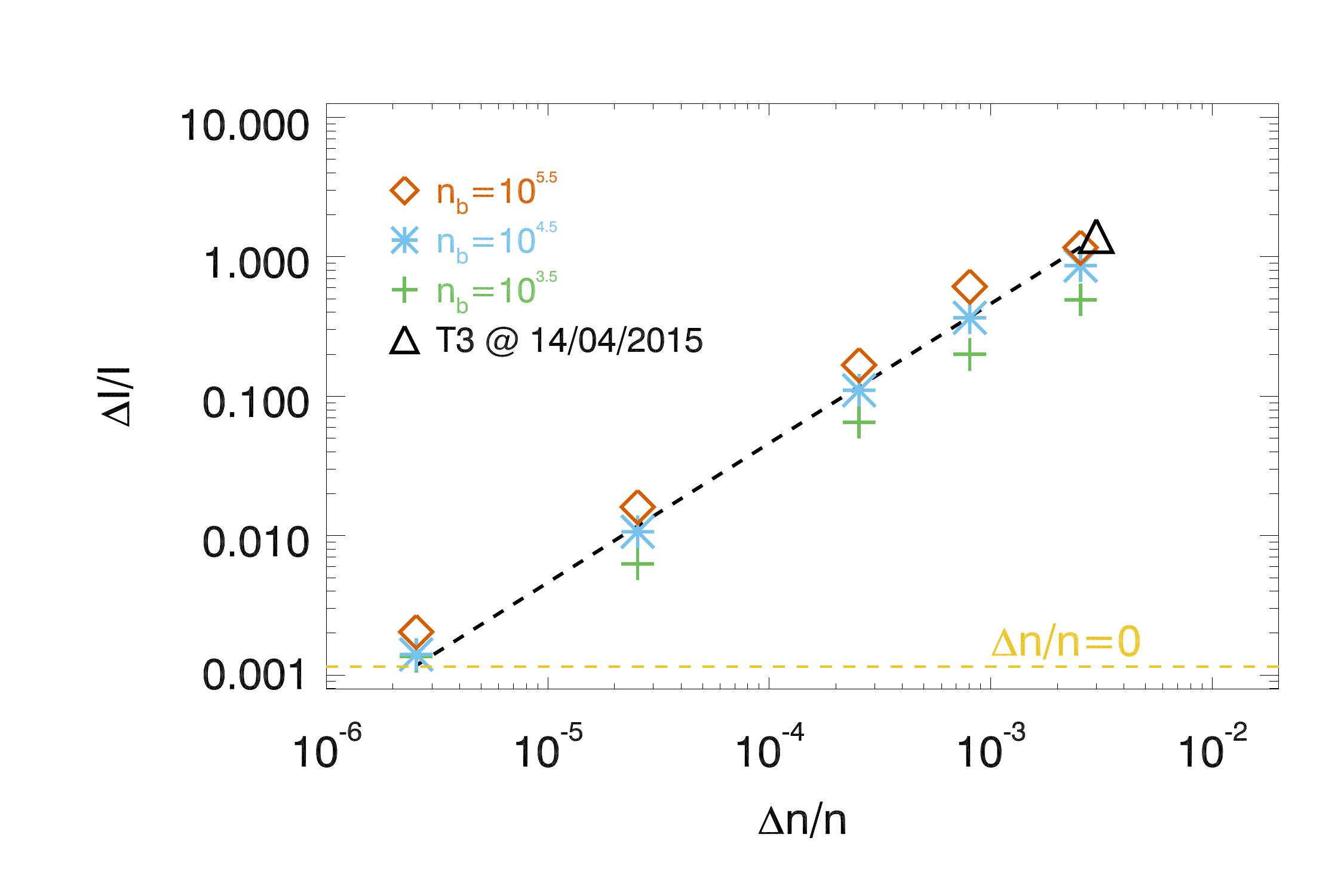}
\caption{\textbf{Relation between the level of background density fluctuations and type III radio burst intensity fluctuations.} Simulated levels of radio emission fluctuations $\Delta I/I$, for electron beams travelling through plasma with varying levels of density fluctuations $\Delta n/n$.  Values were taken between 30--40 MHz from the simulated dynamic spectra.  The case when an electron beam propagates through a homogeneous background plasma is indicated as a yellow dashed line, showing the minimum level of fluctuations obtained by the method.  The initial beam density varied between $10^{3.5},10^{4.5},10^{5.5}~\rm{cm}^{-3}$ indicated in green, blue and red, respectively.  The value of $\Delta I/I$ is shown in black for the type III event observed by LOFAR on the 14/04/2015, at the estimated value of $\Delta n/n$.  The black dashed line shows the fit found using Equation \ref{eqn:didn} from the estimated values of $v_b$ and $v_{th}$ obtained from the type III data.}
\label{figure:sun_dii_dnn}
\end{figure}

The characteristic levels of radio fine structure can now be used as a new remote sensor for density turbulence in the solar corona and solar wind.  Using Equation \ref{eqn:didn} and the derived values for $\Delta I/I, v_b, v_{\rm Th}$ from the LOFAR observations, we can estimate the amplitude of density fluctuations as $\Delta n/n=0.3\%$.  The simulations provide a robust check for Equation \ref{eqn:didn} as they consider additional physical terms.  Using multiple simulations with different values of $\Delta n/n$,   Figure \ref{figure:sun_dii_dnn} shows how $\Delta I/I$, found between 30--40~MHz (Figure \ref{figure:freq_bt}), corresponding to a length of nearly 100 Mm, increases when the intensity of the turbulence, $\Delta n/n$, is increased.  

The simulations find a good agreement with Equation \ref{eqn:didn}, estimating the bulk electron beam velocity from the synthetic dynamic spectra.  We also show in Figure \ref{figure:sun_dii_dnn} the effect of varying the initial beam density.  Electron beams travel at higher velocities for higher beam densities (with the same initial energy spectral index), on account of an increased electron energy density \citep{Reid:2018ab}.  Correspondingly, the value of $\Delta I/I$ is increased.

In contrast to previous studies \citep{Mugundhan:2017aa,Sharykin:2018aa}, the value of $\Delta f/f$ measured from the striae bursts cannot be used to extrapolate the intensity of the background density turbulence $\Delta n/n$.  As we demonstrated in Figure \ref{figure:lwed_fft_sin_200} using a single characteristic wavelength to perturb a background plasma with constant mean density, modifying $\Delta n/n$ does not change the length scale of the Langmuir wave enhancements.  A similar result is demonstrated in Figure \ref{figure:freq_bt}
where the width $\Delta f/f$ of the striae are not significantly altered despite an order of magnitude change in $\Delta n/n$.  We note that the distribution of turbulence scales will affect the resultant value of $\Delta f/f$ and indeed the value of $\Delta I/I$.  However, the relationship between $\Delta f/f$ and $\Delta n/n$ is not simple enough that one can be used to deduce the other.

\begin{figure}
\center
\includegraphics[width=\wfig]{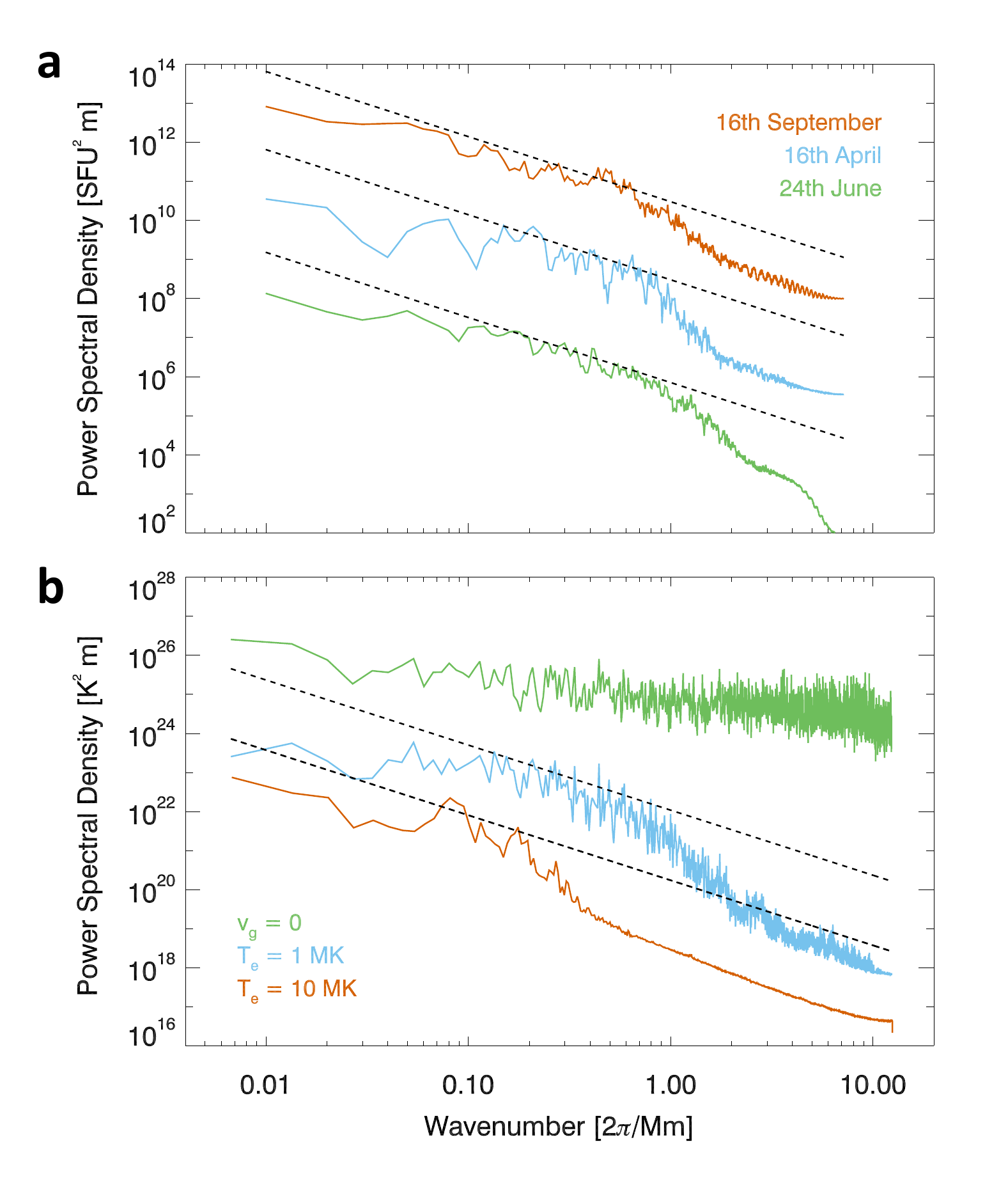}
\caption{\textbf{Power density spectra from observed and simulated type III burst dynamic spectra.} \textbf{a:} power density spectra of the type III bursts as a function of wavenumber assuming the Parker density model.  The 16th September 2015 (red) and the 16th April 2015 (blue) have been shifted up by $10^4$ and $10^2$ for clarity with the 24th June (green).  \textbf{b:} power spectral density for the simulated type III brightness temperature with a 1 MK (blue) and 10 MK (red) corona. Note the reduction in power for the 10 MK corona for higher wavenumbers.  For both graphs, the black dashed lines are a power-law with spectral index -5/3, for reference.  The power spectral density for the simulation with $v_g=0$ is shown in green, multiplied by $10^3$ to better fit into the plot.  The lack of reduction in power at higher wavenumbers (shorter wavelengths) is evident.}
\label{figure:ps_sun_sims}
\end{figure}

Fourier analysis provides knowledge of the pertinent length scales of solar corona and solar wind density turbulence, relevant to the type III fine structure observed by LOFAR (see Methods).  This approach was previously done \citep{Chen:2018aa}, where spectral indices were found around -5/3 for the event on the 16th April in both fundamental and harmonic emission.   Figure \ref{figure:ps_sun_sims} shows all three type III fundamental radio flux power spectra as a function of wavenumber.  At wavenumbers below $2\pi~\rm{Mm}^{-1}$, the spectrum has a roughly -5/3 spectral index.  Around $2\pi~\rm{Mm}^{-1}$, the power in the spectrum decreases, indicating a reduction in the amount of fine structure at the smallest scales.  Figure \ref{figure:ps_sun_sims} also shows the power spectrum from the simulated type III burst in the 2~MK corona.  Similar behaviour is observed to the observations; a spectral index of -5/3 at wavenumbers below $2\pi~\rm{Mm}^{-1}$, and a drop in power around $2\pi~\rm{Mm}^{-1}$.  The drop in power is related to smoothing of Langmuir wave energy from the spatial motion at the group velocity.  Figure \ref{figure:ps_sun_sims} also shows the power spectrum of the type III burst in the 10~MK plasma. With a higher thermal velocity, and hence higher Langmuir wave group velocities, the decrease in power occurs at larger length scales or smaller wavenumbers.  Figure \ref{figure:ps_sun_sims} also shows the power spectrum of the type III bursts where no group velocity was simulated.  The lack of spatial Langmuir wave motion leads to a hightened power at higher wavenumbers, not consisted with the LOFAR observations.

\section{Discussion}

Through a combined observational and simulation approach, we have quantitatively demonstrated how solar radio burst striae fine structure is developed.  The individual radio stria are shown to be a combination
of beam-driven Langmuir wave modulation by the turbulent solar wind, and the subsequent spatial motion of the Langmuir waves at their group velocity.  The simulations indicate that the drift rate of stria is determined by the group velocity of Langmuir waves.

The intensity of the radio fine structure is a powerful tool that can provide a diagnostic of the intensity of the background density turbulence via the relation $\Delta I/I = (v_{\rm Th}^2/v_b^2) \Delta n/n$.  At certain scales, the observed spectrum of radio burst fine structure is very similar to the spectrum of density turbulence measured in the solar wind.  Both the intensity and the spectrum of density turbulence is not well constrained in the solar corona. The LOFAR observations suggest that the level of parallel density fluctuations at heights 0.7-0.8 $R_\odot$ above the photosphere is around $0.1-0.3$\%. This value for $\Delta n/n$ is similar to what was estimated using LOFAR observations of frequency width spread in coronal altitude \citep{Zhang:2019ab}.  The magnitude of parallel density fluctuations is likely smaller than perpendicular density fluctuations, found from anisotropic scattering ray-tracing estimates \citep{Kontar:2019aa}, and similar to values of $\delta B_{\perp}/\delta B_{||}>1$, measured in the solar wind \citep{Chen:2012aa}.  The levels of $\Delta n/n$ are smaller than was previously found \citep{Krupar:2020aa} using type III bursts at lower frequencies, which could be related to their isotropic scattering assumption and density turbulence being stronger in the solar wind.

The frequency drift of radio fine structure constrains the background thermal velocity, increasing the scope of solar radio bursts to be used as a remote plasma temperature diagnostic.  The observation infers a corresponding coronal plasma temperature around $1.1$~MK.  The radio fine structure also provides an additional way to estimate the electron beam bulk velocity which is mostly controlled by the beam energy density.  


Our results have created a framework for exploiting the diagnostic potential of radio burst fine structure.  This is especially relevant given the enhanced resolution of new-age ground-based radio telescopes that are resolving much more fine structure originating from the solar corona.  Moreover, the closer proximity of Parker Solar Probe and Solar Orbiter to radio emission originating in the very high corona or solar wind, and hence higher sensitivity, allows fine structures to be detected in situ.  Coupled with in situ plasma measurements from these spacecraft, 
the radial evolution of the turbulence can be studied throughout the inner heliosphere and should help to understand what drives solar wind turbulence.


\pagebreak

\noindent All correspondence and request for materials should be addressed to HASR at \href{mailto:hamish.reid@ucl.ac.uk}{hamish.reid@ucl.ac.uk}

\section*{Acknowledgements}
We acknowledge support by the Science and Technology Facilities Council (STFC) Consolidated Grant ST/L000533/1.
This work benefited from the Royal Society grant RG130642.  This paper is based (in part) on data obtained with the International LOFAR Telescope (ILT) under project codes LC3\_012 and LC4\_016. LOFAR \citep{van-Haarlem:2013aa} is the Low Frequency Array designed and constructed by ASTRON. It has observing, data processing, and data storage facilities in several countries, that are owned by various parties (each with their own funding sources), and that are collectively operated by the ILT foundation under a joint scientific policy. The ILT resources have benefitted from the following recent major funding sources: CNRS-INSU, Observatoire de Paris and Universit\'e d'Orl\'eans, France; BMBF, MIWF-NRW, MPG, Germany; Science Foundation Ireland (SFI), Department of Business, Enterprise and Innovation (DBEI), Ireland; NWO, The Netherlands; The Science and Technology Facilities Council, UK

\section*{Author contributions}
Both HASR and EPK contributed towards the theoretical, numerical and data analysis required for this study, and for the writing of the text.  Figures were created by HASR.

\section*{Author information}
The authors declare no competing financial interests.  HASR Orchid ID: 0000-0002-6287-3494.  EPK Orchid ID: 0000-0002-8078-0902



\section*{Methods}


\subsection*{Observations}

The LOw Frequency ARray (LOFAR) is a collection of interferometric antenna arrays distributed around Europe, with the core stations in the Netherlands.  Our observations used the tied-array beam forming \citep{Stappers:2011aa} with 24 core stations, using the Low Band Antennas (LBA) between 30-80 MHz.  
Each beam pointed at a different part of the sky and recorded a simultaneous flux that is a convolution of the true source and the LOFAR point spread function.  
We calibrated the flux using an observation of Taurus A (Crab Nebula).  
We use a frequency channel resolution of 0.012 MHz and integrated the 0.01 second time resolution to 0.1 second to increase the signal to noise ratio.  
For imaging, the 124 (April observation) or 169 (June and September observations) individual beams performed a course mosaic of the solar disc and the solar corona, with a separation that is smaller than the FWHM of the LOFAR point spread function.
Images are then made using an interpolation grid.

The drift rate of the type III burst is related to the bulk electron beam velocity via
\begin{equation}
    \pdv{f}{t} = \frac{f}{2n_e}\pdv{n_e}{x}v_b.
\end{equation}
To determine the bulk electron beam velocities from the observations, we converted frequency $\nu$ to background electron density $n_e$ assuming fundamental plasma emission such that $2\pi\nu=\sqrt{4\pi n_e e^2/m_e}$.  
The altitude was then found using the Parker electron density model \citep{Parker:1958aa}.  The bulk electron beam velocity is obtained using a linear fit to the distances as a function of time, 
with a Savisky-Golay filter to smooth the frequency fine structure.

\begin{figure}
\center
\includegraphics[width=\wfig]{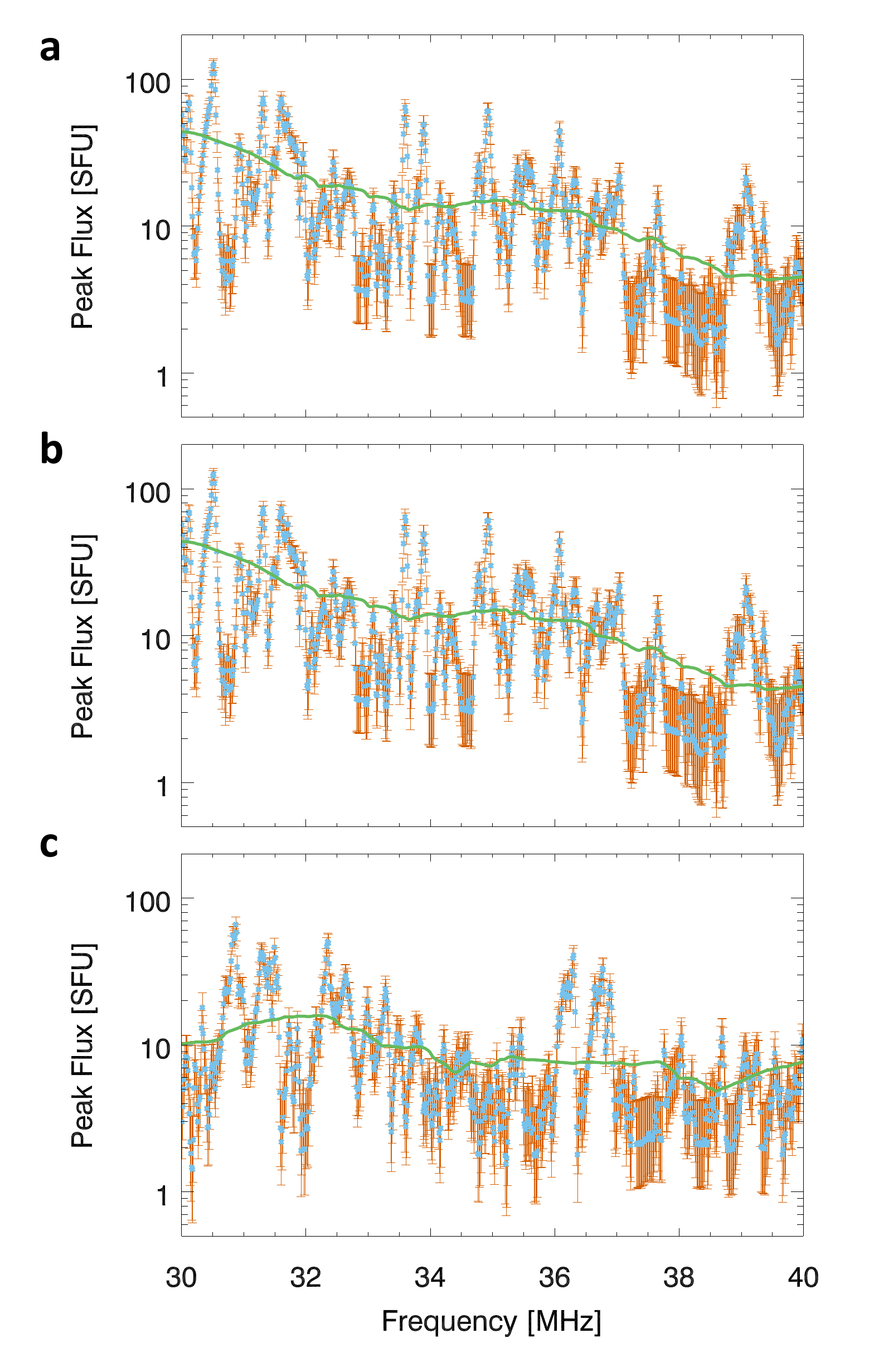}
\caption{\textbf{Illustration of the observed type III radio intensity fluctuations.} The peak flux for the fundamental emission of the three type III bursts observed by LOFAR, shown in Figure \ref{figure:typeIII_dynspec}, on \textbf{a:} 16th April, \textbf{b:} 24th June and \textbf{c:} 16th September.  Errors calculated from the square root of the radio flux are indicated in red.  The smoothed functions, used to calculate the values of $\Delta I/I$ are shown in green.}
\label{figure:typeIII_peakflux}
\end{figure}

The characteristic intensity of the frequency fine structure 
$\Delta I/I$ is obtained using
\begin{equation}
\frac{\Delta I}{I} = \left(\frac{\langle (\delta I(\nu))^2\rangle}{\langle I(\nu)\rangle^2}\right)^{0.5}\,,
\end{equation}
where $I(\nu)$ is the peak flux as a function of frequency and $\delta I(\nu)$ 
is the difference between the peak flux and the smoothed peak flux, found using a Savitsky-Golay filter with a characteristic size of 3 MHz.  
\ref{figure:typeIII_peakflux} 
shows the peak flux and the smoothed peak flux for each event as a function of frequency.

\subsection*{Langmuir wave motion}

Firstly let us note that the decrease of Langmuir wave energy is due to locally increasing density,
i.e. parallel density gradient $dn/dx>0$.  We can write the energy conservation equation \citep{Kontar:2001ab}
\[
E_{w}+E_{b}=const
\]
where $E_{waves}$, $E_{beam}$ are the Langmuir wave and the electron beam energies.

Since the electron beam is almost a plateau $f\simeq n_b/v_0$ up to velocity $v_0$
\[
E_{b}=\int _0^{v_0} mv^2 n_b/v dv
\]
and
\[
E_{w}=\int _0^{v_0}m/\omega_{pe} v^4 (1-v/v_0)n_b/v dv/v^2
\]

Due to the positive density gradient, Langmuir wave packet experiences refraction,
so that the change of phase speed $\delta v$ \citep[see Equations (26, 17) in ][]{Kontar:2001ab} is
\begin{equation}\label{eq:dv_v_kontar}
\frac{\delta v}{v}\simeq \frac{1}{3} \frac{v^2}{v_{Te}^2}\frac{\delta \omega_{pe}}{\omega_{pe}}=\frac{1}{6} \frac{v^2}{v_{Te}^2}\frac{\delta n}{n},
\end{equation}
where $v_{Te}$ is the electron thermal speed.

Since the energy before the refraction and after should be the same,
we can write (energy before = energy after), hence
\[
\frac{mn_b}{6}v_0^2+\frac{mn_b}{12}v_0^2=
\frac{mn_b}{6}(v_0+\delta v)^2+(1-\beta)\frac{mn_b}{12}(v_0+\delta v)^2
\]
the energy from the waves goes to the beam that leads to increase
in the electron energy and must reduce waves by a factor $1-\beta=1-\delta E_w/E_w$,
so that the energy is conserved.

In the limiting case, $1-\beta =0$ (all waves are absorbed), and the maximum
beam velocity is $v_0+\delta v=\sqrt{3/2}$ (e.g. Ryutov, 1969). 
For weak absorption, $\beta << 1$.

From the energy conservation, one finds
\[
3/2v_0^2=(3/2-\beta/2)(v_0+\delta v)^2
\]
which gives us the connection between wave energy change $\beta =\delta E_w/E_w$
and the velocity increase $\delta v/v_0$ becomes
\begin{equation}\label{eq:dv_v}
\beta =3\left(1-(1+\delta v/v_0)^{-2}\right)=3\left(1-(1+1/6 (v^2/v_{Te}^2)\delta n/n)^{-2}\right)
\end{equation}
For small speed change $\delta v/v \ll 1$, one finds
\[
\frac{\delta v}{v_0}\simeq \frac{\beta}{6}=\frac{1}{6}\frac{\delta E_w}{E_w}
\]
Combining the energy conservation and the phase velocity change,
one finds for ${\delta E_w}/{E_w}$:
\begin{equation}\label{eq:deltaE_w}
  \frac{\delta n}{n}=\frac{v_{Te}^2}{v_0^2}\frac{\delta E_w}{E_w}.
\end{equation}

However, if the modulation is known from the observations we can determine
the density fluctuations. Thus taking 50\% intensity modulation of radio flux
\citep[as observed by][]{Kontar:2017aa},
i.e. $\delta E_w/E_w=1/2$, one finds from the Equation~\ref{eq:dv_v}:
\[
\frac{\delta v}{v_0}=\sqrt{6/5}-1\simeq 1/10
\]
then using Equation \ref{eq:dv_v_kontar}, we find:
\[
\frac{\delta n}{n}=\frac{6}{10} \frac{v_{Te}^2}{v^2}
\]
for the solar corona temperature $T=2$~MK, $v_{Te}=\sqrt{k_BT/m}\simeq5.5$~Mm/s and the speed of the electron beam $v=1/3c$, one finds
\[
\frac{\delta n}{n}\simeq 9\times 10^{-4}
\]

It is interesting to compare the spectrum of radio wave intensity modulation
$\langle \delta I^2\rangle _k$ and the spectrum of parallel density fluctuations
$\langle \delta n^2\rangle _k$. From Equation \ref{eq:deltaE_w}, one obtains
the relation
\begin{equation}\label{eq:dn_spectr}
  \frac{\langle \delta n^2\rangle _k}{n^2}=\frac{v_{Te}^4}{v_0^4}\frac{\langle \delta I^2\rangle _k}{I^2}
\end{equation}

Observationally, \citep[from][]{Chen:2018aa},
we have $\langle \delta I^2\rangle _k/I^2=10^{-4}$ at 1~Mm. 
 It is also interesting to note that the expression given by Equation \ref{eq:dv_v} saturates.

\subsection*{Simulation description}


We model the self-consistent time evolution of an electron beam distribution function $f(v,x,t)$ [cm$^{-4}$ s$^{-1}$] and their resonant interaction with the Langmuir waves spectral energy density $W(v,x,t)$ [ergs cm$^{-3}$] with the following 1D kinetic equations
\begin{eqnarray}\label{eqk1}
\frac{\partial f}{\partial t} + \frac{v}{M(r)}\frac{\partial}{\partial r}M(r)f =
\frac{4\pi ^2e^2}{m_e^2}\frac{\partial }{\partial v}\left(\frac{W}{v}\frac{\partial f}{\partial v}\right)
	 +\frac{4\pi n_e e^4}{m_e^2}\ln\Lambda\frac{\partial}{\partial v}\frac{f}{v^2} + S(v,r,t),
\end{eqnarray}
\begin{eqnarray}\label{eqk2}
\frac{\partial W}{\partial t} + \frac{\partial \omega_L}{\partial k}\frac{\partial W}{\partial r}
-\frac{\partial \omega _{pe}}{\partial r}\frac{\partial W}{\partial k}
= \frac{\pi \omega_{pe}}{n_e}v^2W\frac{\partial f}{\partial v}
- (\gamma_{L} +\gamma_c )W + e^2\omega_{pe}v f \ln{\frac{v}{v_{Te}}},
\end{eqnarray}
A complete description of Equations \ref{eqk1} and \ref{eqk2} can be found in previous works \citep[e.g.][]{Reid:2018ab}.  Equation \ref{eqk1} simulates the electron propagation along a guiding magnetic flux rope, together with a decrease in density as the guiding magnetic flux rope expands (modelled through the cross-sectional area $M(r)$ of the expanding flux tube).  Quasilinear terms \citep{Vedenov:1963aa,Drummond:1964aa} in both equations describe the resonant wave growth ($\omega_{pe}=kv$) and the subsequent diffusion of electrons in velocity space.  The background plasma Landau damping rate is $\gamma_L$.  Collisions of electron and waves ($\gamma_c$) modify $f(v,x,t)$ and $W(v,x,t)$ primarily in the dense solar corona.
Equation \ref{eqk2} models the spontaneous emission of waves \citep[e.g.][]{Zheleznyakov:1970aa},
the propagation of waves, and importantly the refraction of waves on density fluctuations \citep[e.g.][]{Ryutov:1969aa}.


We approximate a dynamic spectrum of fundamental emission from the Langmuir wave spectral energy density assuming a saturation level of plasma emission \citep{Melrose:1980aa,Tsytovich:1995aa,Lyubchyk:2017aa}.
The brightness temperature $T_T(k,r,t)$ is found using
\begin{equation} \label{eqn:t_b}
k_b T_T(k,r,t) \approx \frac{(2\pi)^2}{k_L(r)^2}W_L(k,r,t).
\end{equation}
We use the peak value of $T_T(k,r,t)$ to obtain the brightness temperature at each position (frequency), $T_T(r,t)$, as the spread in $k$ is small.

The electron beam is injected as a source function
\begin{equation}\label{eqn:source}
S(v,r,t) = A_v v^{-\alpha}\exp\left(-\frac{r^2}{d^2}\right)A_t\exp\left(-\frac{(t-t_{\rm inj})^2}{\tau^2}\right).
\end{equation}
The velocity distribution is a power-law characterised by $\alpha$, the velocity spectral index.  The constant $A_v\propto n_{\rm beam}$ scales the injected distribution such that the integral over velocity between $v_{\rm min}$ and $v_{\rm max}$ gives the number density $n_{\rm beam}$ of injected electrons.  The spatial distribution is characterised by $d$ [cm], the spread of the electron beam in distance.  The temporal profile is characterised by $\tau$ [s], where $t_{\rm inj}=4\tau$.  It is normalised by $A_t$ such that the integral over time is 1.

The thermal level of Langmuir waves is set to
\begin{equation}\label{eqn:init_w}
W^{\rm init}(v,r,t=0) = \frac{k_BT_e}{4\pi^2}\frac{\omega_{pe}^2}{v^2}\ln\left(\frac{v}{v_{Te}}\right),
\end{equation}
where $k_B$ is the Boltzmann constant and $T_e$ is the electron temperature.  Equation \ref{eqn:init_w} represents the thermal level of spontaneously emitted Langmuir waves from an uniform Maxwellian background plasma when Coulomb collisions are neglected.

We introduce a population of thermal electrons as a background plasma.  This background Maxwellian population is characterised by a background temperature $T_e=2$~MK that corresponds to a background thermal velocity of $v_{Te}=5.5\times10^8~\rm{cm~s}^{-1}$.  The choice of $2$~MK is related to the higher Landau damping that is predicted from the strahl present in the heliosphere \citep[e.g.][]{Maksimovic:2005aa}.

For the background electron density $n_0(r)$, we calculate the smooth background density profile using the Parker model that solves the equations for a stationary spherical symmetric solution \citep{Parker:1958aa} with normalisation factor found from satellites \citep{Mann:1999aa}.
\begin{equation}\label{sol1}
r^2n_e(r)v(r)= C= const
\end{equation}
\begin{equation}\label{sol2}
  \frac{v(r)^2}{v_c^2}-\mbox{ln}\left(\frac{v(r)^2}{v_c^2}\right)=
  4\mbox{ln}\left(\frac{r}{r_c}\right)+4\frac{r_c}{r}-3
\end{equation}
where the critical velocity $v_c$ is defined such that $v_c\equiv (k_BT_{sw}/\tilde{\mu}m_p)^{1/2}$ and the critical radius is defined by $r_c(v_c) = GM_s/2v_c^2$ (both independent on $r$).  $T_{sw}$ is the temperature of the solar wind,\footnote{$T_{sw}$ used in the density model is different from the electron temperature $T_e$ that defines $v_{Te}$} taken as 1 MK, $M_s$ is the mass of the Sun, $m_p$ is the proton mass and $\tilde{\mu}$ is the mean molecular weight. The constant appearing above is fixed by satellite measurements near the Earth's orbit (at $r = 1$~AU, $n =6.59$~cm$^{-3}$) and equates to $6.3\times 10^{34}$~s$^{-1}$.  This model is static in time, set at the start of the simulations, justified through the electron beam moving at least two orders of magnitude faster than the solar wind velocity.

Static background fluctuations are added because the propagating electron beam is travelling much faster than any change in the background density.
The spectrum of density fluctuations has a spectral index $-5/3$, similar to observations near the Earth \citep[e.g.][]{Celnikier:1983aa,Celnikier:1987aa,Chen:2013ab} between wavelengths 1--100~Mm, so that the perturbed density profile is given by
\begin{equation}\label{eqn:fluc}
n_e(r) = n_0(r)\left[1 + C(r)\sum_{n=1}^N\lambda_n^{\mu/2}\sin(2\pi r/\lambda_n + \phi_n)\right]\,,
\end{equation}
where $N=1000$ is the number of perturbations, $n_0(r)$ is the initial unperturbed density as defined above, $\lambda_n$ is the wavelength of the $n$-th fluctuation, $\mu=5/3$ is the power-law spectral index in the power spectrum, and $\phi_n$ is the random phase of the individual fluctuations.  $C(r)$ is the normalisation constant the defines the r.m.s. deviation of the density $\sqrt{\langle \Delta n(r)^2 \rangle}$ such that
 \begin{equation}
C(r) = \sqrt{\frac{2\langle \Delta n(r)^2 \rangle}{\langle n(r) \rangle^2\sum_{n=1}^N\lambda_n^{\mu}}}.
\end{equation}
Our one-dimensional approach means that we are only modelling fluctuations parallel to the magnetic field and not perpendicular.

We model an electron beam injection into the corona with a timescale $\tau=0.001$~s that was near instantaneous to replicate energisation via magnetic reconnection.  We used a reasonably dilute electron beam where $n_b/n_e=10^{-5}$, accelerated in a region of longitudinal length 10~Mm, typical for coronal electron beams \citep{Reid:2011aa}.
The electrons initially have a power-law distribution with a spectral index
of 8 in velocity space, ranging from $4.4-38~v_{\rm th}$ ($1.7-125$~keV).  Electron populations with smaller or larger velocities typically do not produce significant levels of Langmuir waves.

The electron density is exponential in the solar corona \citep[see][for details]{Parker:1958aa}, and has approximately $r^{-2}$ decrease in interplanetary space.  Additionally, turbulence over a spectrum of length scales typically below $100$~Mm fluctuates the electron density with an intensity that increases with distance from the solar surface \citep{Woo:1995aa,Mugundhan:2017aa,Sasikumar-Raja:2017aa,Krupar:2018aa}, but remains much more constant over interplanetary distances \cite{Bisoi:2014aa}.  We modelled the turbulence by increase the turbulent intensity $\Delta n/n$ as a function of distance, reaching a value of $\Delta n/n = 10^{-1.5}$ near the Earth.
The coronal plasma had a bulk temperature of $1$~MK,
giving a thermal velocity of $3.9~\rm{Mm~s}^{-1}$.

\subsection*{Density fluctuation wavelength}

The simulations carried out in Section \ref{sec:lwaves} were very similar to the main type III burst simulations but with some notable changes.  The background density $n_0(r)$ was set constant to $1.1\times10^{7}~\rm{cm}^{-3}$ that corresponds to a plasma frequency of 30 MHz.  The initial beam density was set as a broken power-law in velocity space, varying as $v^{-\alpha}$ when $v>v_0$, with a break energy of 10 keV, and constant when $v<v_0$.  The reduction is electron flux at lower velocities was to account for the lack of Coulomb collisions in the lower corona.  The second term in Equation \ref{eqk1} did not include the radial expansion and such was $v\frac{\partial f}{\partial{r}}$, to enhance the level of Langmuir waves generated by the simulation.    

\begin{figure}
\center
\includegraphics[width=\wfigure]{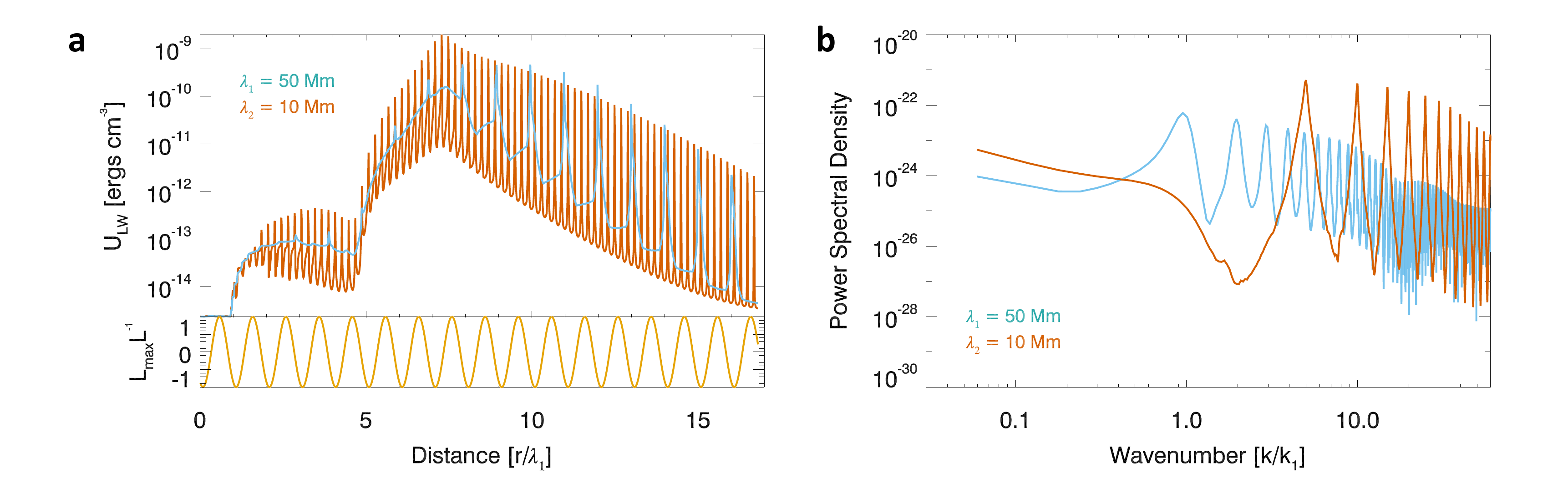}
\caption{\textbf{Simulated beam-drive Langmuir wave energy density and associated power density spectra.}  \textbf{a:}  Beam-generated Langmuir wave energy density $U_{LW}$ as a function of distance.  The background plasma has a sinusoidal perturbation with amplitude $A=0.001$ and wavelength $\lambda=50$~Mm in blue and $\lambda=10$~Mm in red.  The lower panel highlights the modulation in the background plasma gradient length scale $L=0.5n_e^{-1}dn_e/dx$.  \textbf{b:}
Power density spectrum of the Langmuir wave energy density minus the unperturbed case from the simulations shown in panel \textbf{a}.  The peaks occur at the wavenumber $k_1=2\pi/\lambda~\rm{Mm}^{-1}$ and the harmonics.  The simulation with a lower wavelength density fluctuations clearly modulates the Langmuir waves at higher wavenumbers.}
\label{figure:sin_sm}
\end{figure}

The wavelength of background density fluctuations controls how regular the modulations occurs of beam-driven Langmuir waves.  When the wavelength is reduced by a factor of five, from $\lambda_1=50$~Mm to $\lambda_2=10$~Mm, we can see modulation in Langmuir wave growth causes the energy density to oscillate with a correspondingly shorter length scale.  This is shown in \ref{figure:sin_sm}
.  The oscillation causes a peak at higher wavenumbers, shown in Fourier space.

\subsection*{Radio fluctuations}

\begin{figure}
\center
\includegraphics[width=\wfig]{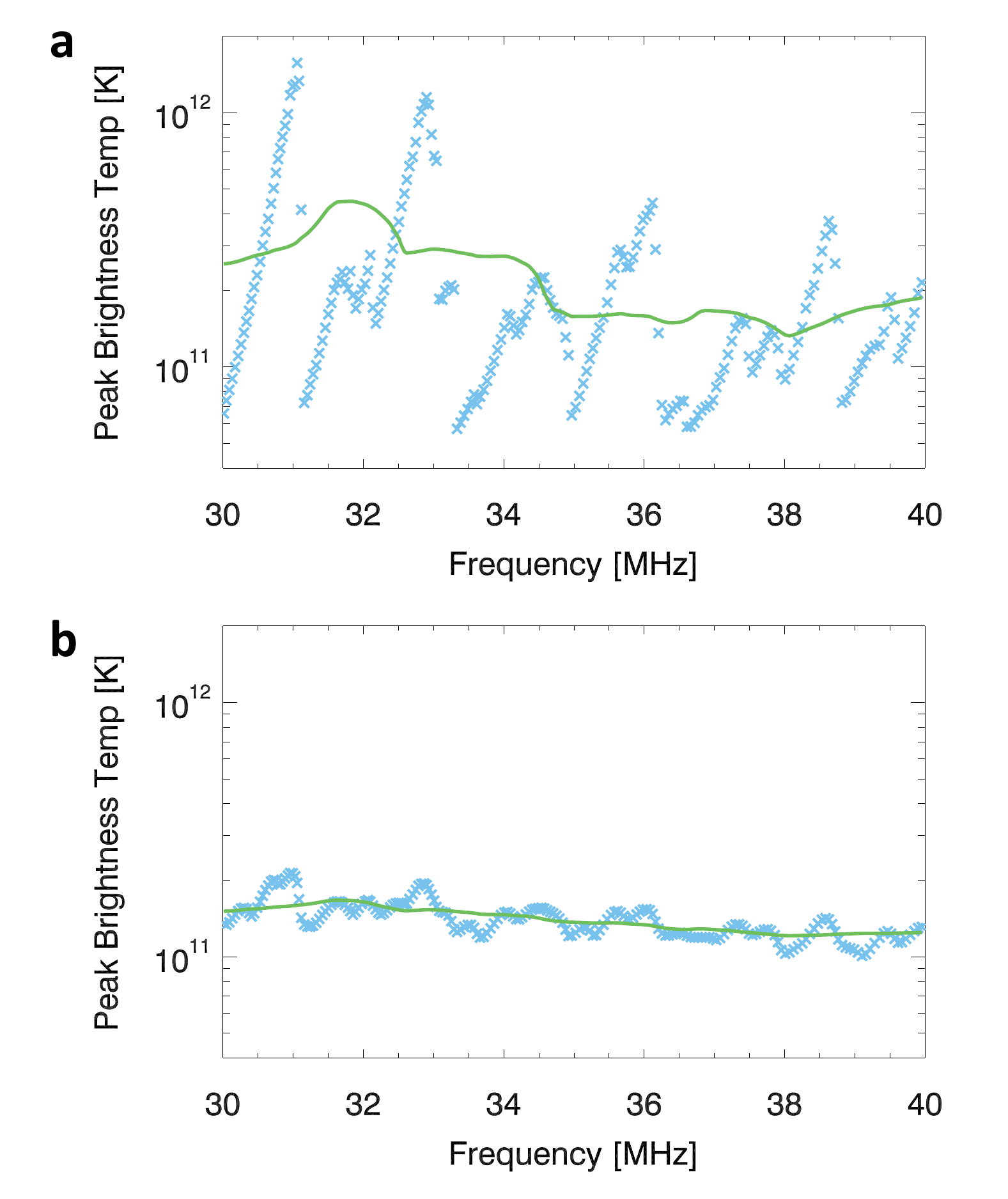}
\caption{\textbf{Illustration of the simulated type III radio intensity fluctuations.} \textbf{a:} peak brightness temperature as a function of frequency for the simulated type III dynamic spectra propagating through plasma where $\Delta n/n$ $2.5\times10^{-3}$.  The smoothed function used to calculate $\Delta I/I$ is overplotted in green.  \textbf{b:} same as panel \textbf{a} but $\Delta n/n$ is $2.5\times10^{-4}$.}
\label{figure:freq_bt}
\end{figure}

We calculate the intensity of the fluctuations in the peak brightness temperature, obtaining $\Delta I/I$, between 30--40~MHz, corresponding to a length of nearly $100$~Mm.  We subtract a smoothed function with a smoothing box of one third the box size, $30$~Mm to find the r.m.s. deviation.  An example of the radio fluctuations and the smoothed function is given in Figure \ref{figure:freq_bt}. 

Comparing the simulated peak brightness temperature with the observed type III peak flux, the striae clumps display a similar pattern of steep rise and shallow decay.  The clump structures in the curves of \ref{figure:typeIII_peakflux}
display a similar pattern to the clump structures in the simulated curves in Figure \ref{figure:freq_bt}
, increasing sharply and then decreasing.  The rise and decay of the radio bursts is not as smooth as the simulated bursts, presumably related to the smaller length scales that are present in the solar corona that were not simulated.

To calculate the power density spectra, we summed the square of the Fourier transform of the type III flux over the duration of the radio bursts between 30-40 MHz.  The corresponding altitude associated with each frequency was found using the Parker density model.  For the simulated radio bursts, we found the Fourier transform between 30-80 MHz to reduce the noise in the power spectrum.



\bibliographystyle{unsrt}
\bibliography{ubib}

\begin{thebibliography}{10}

\bibitem{Suzuki:1985aa}
S.~{Suzuki} and G.~A. {Dulk}.
\newblock {\em {Bursts of Type III and Type V}}, pages 289--332.
\newblock Cambridge University Press, 1985.

\bibitem{Pick:2008aa}
M.~{Pick} and N.~{Vilmer}.
\newblock {Sixty-five years of solar radioastronomy: flares, coronal mass
  ejections and Sun Earth connection}.
\newblock {\em \aapr}, 16:1--153, October 2008.

\bibitem{Reid:2014aa}
H.~A.~S. {Reid}, N.~{Vilmer}, and E.~P. {Kontar}.
\newblock {The low-high-low trend of type III radio burst starting frequencies
  and solar flare hard X-rays}.
\newblock {\em \aap}, 567:A85, July 2014.

\bibitem{Aschwanden:2005aa}
Markus~J. {Aschwanden}.
\newblock {\em {Physics of the Solar Corona. An Introduction with Problems and
  Solutions (2nd edition)}}.
\newblock Praxis Publishing Ltd, 2005.

\bibitem{Holman:2011aa}
G.~D. {Holman}, M.~J. {Aschwanden}, H.~{Aurass}, M.~{Battaglia}, P.~C.
  {Grigis}, E.~P. {Kontar}, W.~{Liu}, P.~{Saint-Hilaire}, and V.~V. {Zharkova}.
\newblock {Implications of X-ray Observations for Electron Acceleration and
  Propagation in Solar Flares}.
\newblock {\em \ssr}, 159:107--166, September 2011.

\bibitem{Dulk:1985aa}
G.~A. {Dulk}.
\newblock {Radio emission from the sun and stars}.
\newblock {\em \araa}, 23:169--224, 1985.

\bibitem{Posner:2007aa}
Arik {Posner}.
\newblock {Up to 1-hour forecasting of radiation hazards from solar energetic
  ion events with relativistic electrons}.
\newblock {\em Space Weather}, 5(5):05001, May 2007.

\bibitem{de-La-Noe:1972aa}
J.~{de La Noe} and A.~{Boischot}.
\newblock {The Type III B Burst}.
\newblock {\em \aap}, 20:55, August 1972.

\bibitem{Abranin:1978aa}
E.~P. {Abranin}, L.~L. {Bazelian}, N.~I. {Goncharov}, V.~V. {Zaitsev}, V.~A.
  {Zinichev}, V.~O. {Rapoport}, and I.~G. {Tsybko}.
\newblock {Angular sizes of stria-burst sources in the range 24-26 MHz}.
\newblock {\em \solphys}, 57:229--235, March 1978.

\bibitem{Kontar:2017aa}
E.~P. {Kontar}, S.~{Yu}, A.~A. {Kuznetsov}, A.~G. {Emslie}, B.~{Alcock},
  N.~L.~S. {Jeffrey}, V.~N. {Melnik}, N.~H. {Bian}, and P.~{Subramanian}.
\newblock {Imaging spectroscopy of solar radio burst fine structures}.
\newblock {\em Nature Communications}, 8:1515, November 2017.

\bibitem{MelNik:2018aa}
V.~N. {Melnik}, A.~I. {Brazhenko}, A.~V. {Frantsuzenko}, V.~V. {Dorovskyy}, and
  H.~O. {Rucker}.
\newblock {Properties of Decameter IIIb-III Pairs}.
\newblock {\em \solphys}, 293:26, February 2018.

\bibitem{Melrose:1983aa}
D.~B. {Melrose}.
\newblock {Frequency splitting in stria bursts - Possible roles of
  low-frequency waves}.
\newblock {\em \solphys}, 87:359--371, September 1983.

\bibitem{Kolotkov:2018aa}
Dmitrii~Y. {Kolotkov}, Valery~M. {Nakariakov}, and Eduard~P. {Kontar}.
\newblock {Origin of the Modulation of the Radio Emission from the Solar Corona
  by a Fast Magnetoacoustic Wave}.
\newblock {\em \apj}, 861(1):33, July 2018.

\bibitem{Sharykin:2018aa}
I.~N. {Sharykin}, E.~P. {Kontar}, and A.~A. {Kuznetsov}.
\newblock {LOFAR Observations of Fine Spectral Structure Dynamics in Type IIIb
  Radio Bursts}.
\newblock {\em \solphys}, 293(8):115, August 2018.

\bibitem{Pecseli:2012aa}
Hans {P{\'e}cseli}.
\newblock {\em {Waves and Oscillations in Plasmas}}.
\newblock Taylor \& Francis, Boca Raton, 2012.

\bibitem{Chen:2018aa}
Xingyao {Chen}, Eduard~P. {Kontar}, Sijie {Yu}, Yihua {Yan}, Jing {Huang}, and
  Baolin {Tan}.
\newblock {Fine Structures of Solar Radio Type III Bursts and Their Possible
  Relationship with Coronal Density Turbulence}.
\newblock {\em \apj}, 856(1):73, March 2018.

\bibitem{Parker:1958aa}
E.~N. {Parker}.
\newblock {Dynamics of the Interplanetary Gas and Magnetic Fields.}
\newblock {\em \apj}, 128:664, November 1958.

\bibitem{van-Haarlem:2013aa}
M.~P. {van Haarlem}, M.~W. {Wise}, A.~W. {Gunst}, G.~{Heald}, J.~P. {McKean},
  J.~W.~T. {Hessels}, A.~G. {de Bruyn}, R.~{Nijboer}, J.~{Swinbank},
  R.~{Fallows}, M.~{Brentjens}, A.~{Nelles}, R.~{Beck}, H.~{Falcke},
  R.~{Fender}, J.~{H{\"o}randel}, L.~V.~E. {Koopmans}, G.~{Mann}, G.~{Miley},
  H.~{R{\"o}ttgering}, B.~W. {Stappers}, R.~A.~M.~J. {Wijers}, S.~{Zaroubi},
  M.~{van den Akker}, A.~{Alexov}, J.~{Anderson}, K.~{Anderson}, A.~{van
  Ardenne}, M.~{Arts}, A.~{Asgekar}, I.~M. {Avruch}, F.~{Batejat},
  L.~{B{\"a}hren}, M.~E. {Bell}, M.~R. {Bell}, I.~{van Bemmel}, P.~{Bennema},
  M.~J. {Bentum}, G.~{Bernardi}, P.~{Best}, L.~{B{\^\i}rzan}, A.~{Bonafede},
  A.-J. {Boonstra}, R.~{Braun}, J.~{Bregman}, F.~{Breitling}, R.~H. {van de
  Brink}, J.~{Broderick}, P.~C. {Broekema}, W.~N. {Brouw}, M.~{Br{\"u}ggen},
  H.~R. {Butcher}, W.~{van Cappellen}, B.~{Ciardi}, T.~{Coenen}, J.~{Conway},
  A.~{Coolen}, A.~{Corstanje}, S.~{Damstra}, O.~{Davies}, A.~T. {Deller}, R.-J.
  {Dettmar}, G.~{van Diepen}, K.~{Dijkstra}, P.~{Donker}, A.~{Doorduin},
  J.~{Dromer}, M.~{Drost}, A.~{van Duin}, J.~{Eisl{\"o}ffel}, J.~{van Enst},
  C.~{Ferrari}, W.~{Frieswijk}, H.~{Gankema}, M.~A. {Garrett}, F.~{de
  Gasperin}, M.~{Gerbers}, E.~{de Geus}, J.-M. {Grie{\ss}meier}, T.~{Grit},
  P.~{Gruppen}, J.~P. {Hamaker}, T.~{Hassall}, M.~{Hoeft}, H.~A. {Holties},
  A.~{Horneffer}, A.~{van der Horst}, A.~{van Houwelingen}, A.~{Huijgen},
  M.~{Iacobelli}, H.~{Intema}, N.~{Jackson}, V.~{Jelic}, A.~{de Jong},
  E.~{Juette}, D.~{Kant}, A.~{Karastergiou}, A.~{Koers}, H.~{Kollen}, V.~I.
  {Kondratiev}, E.~{Kooistra}, Y.~{Koopman}, A.~{Koster}, M.~{Kuniyoshi},
  M.~{Kramer}, G.~{Kuper}, P.~{Lambropoulos}, C.~{Law}, J.~{van Leeuwen},
  J.~{Lemaitre}, M.~{Loose}, P.~{Maat}, G.~{Macario}, S.~{Markoff},
  J.~{Masters}, R.~A. {McFadden}, D.~{McKay-Bukowski}, H.~{Meijering},
  H.~{Meulman}, M.~{Mevius}, E.~{Middelberg}, R.~{Millenaar}, J.~C.~A.
  {Miller-Jones}, R.~N. {Mohan}, J.~D. {Mol}, J.~{Morawietz}, R.~{Morganti},
  D.~D. {Mulcahy}, E.~{Mulder}, H.~{Munk}, L.~{Nieuwenhuis}, R.~{van
  Nieuwpoort}, J.~E. {Noordam}, M.~{Norden}, A.~{Noutsos}, A.~R. {Offringa},
  H.~{Olofsson}, A.~{Omar}, E.~{Orr{\'u}}, R.~{Overeem}, H.~{Paas},
  M.~{Pandey-Pommier}, V.~N. {Pandey}, R.~{Pizzo}, A.~{Polatidis},
  D.~{Rafferty}, S.~{Rawlings}, W.~{Reich}, J.-P. {de Reijer}, J.~{Reitsma},
  G.~A. {Renting}, P.~{Riemers}, E.~{Rol}, J.~W. {Romein}, J.~{Roosjen},
  M.~{Ruiter}, A.~{Scaife}, K.~{van der Schaaf}, B.~{Scheers}, P.~{Schellart},
  A.~{Schoenmakers}, G.~{Schoonderbeek}, M.~{Serylak}, A.~{Shulevski},
  J.~{Sluman}, O.~{Smirnov}, C.~{Sobey}, H.~{Spreeuw}, M.~{Steinmetz}, C.~G.~M.
  {Sterks}, H.-J. {Stiepel}, K.~{Stuurwold}, M.~{Tagger}, Y.~{Tang},
  C.~{Tasse}, I.~{Thomas}, S.~{Thoudam}, M.~C. {Toribio}, B.~{van der Tol},
  O.~{Usov}, M.~{van Veelen}, A.-J. {van der Veen}, S.~{ter Veen}, J.~P.~W.
  {Verbiest}, R.~{Vermeulen}, N.~{Vermaas}, C.~{Vocks}, C.~{Vogt}, M.~{de Vos},
  E.~{van der Wal}, R.~{van Weeren}, H.~{Weggemans}, P.~{Weltevrede},
  S.~{White}, S.~J. {Wijnholds}, T.~{Wilhelmsson}, O.~{Wucknitz},
  S.~{Yatawatta}, P.~{Zarka}, A.~{Zensus}, and J.~{van Zwieten}.
\newblock {LOFAR: The LOw-Frequency ARray}.
\newblock {\em \aap}, 556:A2, August 2013.

\bibitem{Lemen:2012aa}
J.~R. {Lemen}, A.~M. {Title}, D.~J. {Akin}, P.~F. {Boerner}, C.~{Chou}, J.~F.
  {Drake}, D.~W. {Duncan}, C.~G. {Edwards}, F.~M. {Friedlaender}, G.~F.
  {Heyman}, N.~E. {Hurlburt}, N.~L. {Katz}, G.~D. {Kushner}, M.~{Levay}, R.~W.
  {Lindgren}, D.~P. {Mathur}, E.~L. {McFeaters}, S.~{Mitchell}, R.~A. {Rehse},
  C.~J. {Schrijver}, L.~A. {Springer}, R.~A. {Stern}, T.~D. {Tarbell}, J.-P.
  {Wuelser}, C.~J. {Wolfson}, C.~{Yanari}, J.~A. {Bookbinder}, P.~N.
  {Cheimets}, D.~{Caldwell}, E.~E. {Deluca}, R.~{Gates}, L.~{Golub}, S.~{Park},
  W.~A. {Podgorski}, R.~I. {Bush}, P.~H. {Scherrer}, M.~A. {Gummin},
  P.~{Smith}, G.~{Auker}, P.~{Jerram}, P.~{Pool}, R.~{Soufli}, D.~L. {Windt},
  S.~{Beardsley}, M.~{Clapp}, J.~{Lang}, and N.~{Waltham}.
\newblock {The Atmospheric Imaging Assembly (AIA) on the Solar Dynamics
  Observatory (SDO)}.
\newblock {\em \solphys}, 275:17--40, January 2012.

\bibitem{Reid:2010aa}
H.~A.~S. {Reid} and E.~P. {Kontar}.
\newblock {Solar Wind Density Turbulence and Solar Flare Electron Transport
  from the Sun to the Earth}.
\newblock {\em \apj}, 721:864--874, September 2010.

\bibitem{Li:2012aa}
B.~{Li}, I.~H. {Cairns}, and P.~A. {Robinson}.
\newblock {Frequency Fine Structures of Type III Bursts Due to Localized
  Medium-Scale Density Structures Along Paths of Type III Beams}.
\newblock {\em \solphys}, 279:173--196, July 2012.

\bibitem{Loi:2014aa}
S.~T. {Loi}, I.~H. {Cairns}, and B.~{Li}.
\newblock {Production of Fine Structures in Type III Solar Radio Bursts Due to
  Turbulent Density Profiles}.
\newblock {\em \apj}, 790:67, July 2014.

\bibitem{Reid:2017ab}
H.~A.~S. {Reid} and E.~P. {Kontar}.
\newblock {Langmuir wave electric fields induced by electron beams in the
  heliosphere}.
\newblock {\em \aap}, 598:A44, January 2017.

\bibitem{Melrose:1986aa}
D.~B. {Melrose}, I.~H. {Cairns}, and G.~A. {Dulk}.
\newblock {Clumpy Langmuir waves in type III solar radio bursts}.
\newblock {\em \aap}, 163:229--238, July 1986.

\bibitem{Reid:2018ab}
H.~A.~S. {Reid} and E.~P. {Kontar}.
\newblock {Spatial Expansion and Speeds of Type III Electron Beam Sources in
  the Solar Corona}.
\newblock {\em \apj}, 867:158, November 2018.

\bibitem{Li:2014aa}
B.~{Li} and I.~H. {Cairns}.
\newblock {Fundamental Emission of Type III Bursts Produced in Non-Maxwellian
  Coronal Plasmas with Kappa-Distributed Background Particles}.
\newblock {\em \solphys}, 289:951--976, March 2014.

\bibitem{Li:2013aa}
B.~{Li} and I.~H. {Cairns}.
\newblock {Type III bursts produced by power law injected electrons in
  Maxwellian background coronal plasmas}.
\newblock {\em Journal of Geophysical Research (Space Physics)},
  118:4748--4759, August 2013.

\bibitem{Newkirk:1961aa}
G.~{Newkirk}, Jr.
\newblock {The Solar Corona in Active Regions and the Thermal Origin of the
  Slowly Varying Component of Solar Radio Radiation.}
\newblock {\em \apj}, 133:983, May 1961.

\bibitem{Mugundhan:2017aa}
V.~{Mugundhan}, K.~{Hariharan}, and R.~{Ramesh}.
\newblock {Solar Type IIIb Radio Bursts as Tracers for Electron Density
  Fluctuations in the Corona}.
\newblock {\em \solphys}, 292(11):155, November 2017.

\bibitem{Zhang:2019ab}
PeiJin {Zhang}, SiJie {Yu}, Eduard~P. {Kontar}, and ChuanBing {Wang}.
\newblock {On the Source Position and Duration of a Solar Type III Radio Burst
  Observed by LOFAR}.
\newblock {\em \apj}, 885(2):140, November 2019.

\bibitem{Kontar:2019aa}
Eduard~P. {Kontar}, Xingyao {Chen}, Nicolina {Chrysaphi}, Natasha L.~S.
  {Jeffrey}, A.~Gordon {Emslie}, Vratislav {Krupar}, Milan {Maksimovic}, Mykola
  {Gordovskyy}, and Philippa~K. {Browning}.
\newblock {Anisotropic Radio-wave Scattering and the Interpretation of Solar
  Radio Emission Observations}.
\newblock {\em \apj}, 884(2):122, October 2019.

\bibitem{Chen:2012aa}
C.~H.~K. {Chen}, A.~{Mallet}, A.~A. {Schekochihin}, T.~S. {Horbury}, R.~T.
  {Wicks}, and S.~D. {Bale}.
\newblock {Three-dimensional Structure of Solar Wind Turbulence}.
\newblock {\em \apj}, 758(2):120, October 2012.

\bibitem{Krupar:2020aa}
Vratislav {Krupar}, Adam {Szabo}, Milan {Maksimovic}, Oksana {Kruparova},
  Eduard~P. {Kontar}, Laura~A. {Balmaceda}, Xavier {Bonnin}, Stuart~D. {Bale},
  Marc {Pulupa}, David~M. {Malaspina}, John~W. {Bonnell}, Peter~R. {Harvey},
  Keith {Goetz}, Thierry {Dudok de Wit}, Robert~J. {MacDowall}, Justin~C.
  {Kasper}, Anthony~W. {Case}, Kelly~E. {Korreck}, Davin~E. {Larson}, Roberto
  {Livi}, Michael~L. {Stevens}, Phyllis~L. {Whittlesey}, and Alexander~M.
  {Hegedus}.
\newblock {Density Fluctuations in the Solar Wind Based on Type III Radio
  Bursts Observed by Parker Solar Probe}.
\newblock {\em \apjs}, 246(2):57, February 2020.

\bibitem{Stappers:2011aa}
B.~W. {Stappers}, J.~W.~T. {Hessels}, A.~{Alexov}, K.~{Anderson}, T.~{Coenen},
  T.~{Hassall}, A.~{Karastergiou}, V.~I. {Kondratiev}, M.~{Kramer}, J.~{van
  Leeuwen}, J.~D. {Mol}, A.~{Noutsos}, J.~W. {Romein}, P.~{Weltevrede},
  R.~{Fender}, R.~A.~M.~J. {Wijers}, L.~{B{\"a}hren}, M.~E. {Bell},
  J.~{Broderick}, E.~J. {Daw}, V.~S. {Dhillon}, J.~{Eisl{\"o}ffel},
  H.~{Falcke}, J.~{Griessmeier}, C.~{Law}, S.~{Markoff}, J.~C.~A.
  {Miller-Jones}, B.~{Scheers}, H.~{Spreeuw}, J.~{Swinbank}, S.~{Ter Veen},
  M.~W. {Wise}, O.~{Wucknitz}, P.~{Zarka}, J.~{Anderson}, A.~{Asgekar}, I.~M.
  {Avruch}, R.~{Beck}, P.~{Bennema}, M.~J. {Bentum}, P.~{Best}, J.~{Bregman},
  M.~{Brentjens}, R.~H. {van de Brink}, P.~C. {Broekema}, W.~N. {Brouw},
  M.~{Br{\"u}ggen}, A.~G. {de Bruyn}, H.~R. {Butcher}, B.~{Ciardi},
  J.~{Conway}, R.-J. {Dettmar}, A.~{van Duin}, J.~{van Enst}, M.~{Garrett},
  M.~{Gerbers}, T.~{Grit}, A.~{Gunst}, M.~P. {van Haarlem}, J.~P. {Hamaker},
  G.~{Heald}, M.~{Hoeft}, H.~{Holties}, A.~{Horneffer}, L.~V.~E. {Koopmans},
  G.~{Kuper}, M.~{Loose}, P.~{Maat}, D.~{McKay-Bukowski}, J.~P. {McKean},
  G.~{Miley}, R.~{Morganti}, R.~{Nijboer}, J.~E. {Noordam}, M.~{Norden},
  H.~{Olofsson}, M.~{Pandey-Pommier}, A.~{Polatidis}, W.~{Reich},
  H.~{R{\"o}ttgering}, A.~{Schoenmakers}, J.~{Sluman}, O.~{Smirnov},
  M.~{Steinmetz}, C.~G.~M. {Sterks}, M.~{Tagger}, Y.~{Tang}, R.~{Vermeulen},
  N.~{Vermaas}, C.~{Vogt}, M.~{de Vos}, S.~J. {Wijnholds}, S.~{Yatawatta}, and
  A.~{Zensus}.
\newblock {Observing pulsars and fast transients with LOFAR}.
\newblock {\em \aap}, 530:A80, June 2011.

\bibitem{Kontar:2001ab}
E.~P. {Kontar}.
\newblock {Dynamics of electron beams in the inhomogeneous solar corona
  plasma}.
\newblock {\em \solphys}, 202:131--149, August 2001.

\bibitem{Vedenov:1963aa}
A.~A. {Vedenov}.
\newblock {Quasi-linear plasma theory (theory of a weakly turbulent plasma)}.
\newblock {\em Journal of Nuclear Energy}, 5:169--186, January 1963.

\bibitem{Drummond:1964aa}
W.~E. {Drummond} and D.~{Pines}.
\newblock {Nonlinear plasma oscillations}.
\newblock {\em Annals of Physics}, 28:478--499, July 1964.

\bibitem{Zheleznyakov:1970aa}
V.~V. {Zheleznyakov} and V.~V. {Zaitsev}.
\newblock {Contribution to the Theory of Type III Solar Radio Bursts. I.}
\newblock {\em \sovast}, 14:47, August 1970.

\bibitem{Ryutov:1969aa}
D.~D. {Ryutov}.
\newblock {Quasilinear Relaxation of an Electron Beam in an Inhomogeneous
  Plasma}.
\newblock {\em Soviet Journal of Experimental and Theoretical Physics}, 30:131,
  1969.

\bibitem{Melrose:1980aa}
D.~B. {Melrose}.
\newblock {\em {Plasma astrohysics. Nonthermal processes in diffuse magnetized
  plasmas - Vol.1: The emission, absorption and transfer of waves in plasmas;
  Vol.2: Astrophysical applications}}.
\newblock Gordon and Breach Science Publishers, 1980.

\bibitem{Tsytovich:1995aa}
V.~N. {Tsytovich} and D.~{ter Haar}.
\newblock {\em {Lectures on Non-linear Plasma Kinetics}}.
\newblock Springer Verlag, 1995.

\bibitem{Lyubchyk:2017aa}
O.~{Lyubchyk}, E.~P. {Kontar}, Y.~M. {Voitenko}, N.~H. {Bian}, and D.~B.
  {Melrose}.
\newblock {Solar Plasma Radio Emission in the Presence of Imbalanced Turbulence
  of Kinetic-Scale Alfv{\'e}n Waves}.
\newblock {\em \solphys}, 292:117, September 2017.

\bibitem{Maksimovic:2005aa}
M.~{Maksimovic}, I.~{Zouganelis}, J.-Y. {Chaufray}, K.~{Issautier}, E.~E.
  {Scime}, J.~E. {Littleton}, E.~{Marsch}, D.~J. {McComas}, C.~{Salem}, R.~P.
  {Lin}, and H.~{Elliott}.
\newblock {Radial evolution of the electron distribution functions in the fast
  solar wind between 0.3 and 1.5 AU}.
\newblock {\em Journal of Geophysical Research (Space Physics)}, 110:A09104,
  September 2005.

\bibitem{Mann:1999aa}
G.~{Mann}, F.~{Jansen}, R.~J. {MacDowall}, M.~L. {Kaiser}, and R.~G. {Stone}.
\newblock {A heliospheric density model and type III radio bursts}.
\newblock {\em \aap}, 348:614--620, August 1999.

\bibitem{Celnikier:1983aa}
L.~M. {Celnikier}, C.~C. {Harvey}, R.~{Jegou}, P.~{Moricet}, and M.~{Kemp}.
\newblock {A determination of the electron density fluctuation spectrum in the
  solar wind, using the ISEE propagation experiment}.
\newblock {\em \aap}, 126:293--298, October 1983.

\bibitem{Celnikier:1987aa}
L.~M. {Celnikier}, L.~{Muschietti}, and M.~V. {Goldman}.
\newblock {Aspects of interplanetary plasma turbulence}.
\newblock {\em \aap}, 181:138--154, July 1987.

\bibitem{Chen:2013ab}
C.~H.~K. {Chen}, G.~G. {Howes}, J.~W. {Bonnell}, F.~S. {Mozer}, K.~G. {Klein},
  and S.~D. {Bale}.
\newblock {Kinetic scale density fluctuations in the solar wind}.
\newblock In G.~P. {Zank}, J.~{Borovsky}, R.~{Bruno}, J.~{Cirtain},
  S.~{Cranmer}, H.~{Elliott}, J.~{Giacalone}, W.~{Gonzalez}, G.~{Li},
  E.~{Marsch}, E.~{Moebius}, N.~{Pogorelov}, J.~{Spann}, and
  O.~{Verkhoglyadova}, editors, {\em American Institute of Physics Conference
  Series}, volume 1539 of {\em American Institute of Physics Conference
  Series}, pages 143--146, June 2013.

\bibitem{Reid:2011aa}
H.~A.~S. {Reid}, N.~{Vilmer}, and E.~P. {Kontar}.
\newblock {Characteristics of the flare acceleration region derived from
  simultaneous hard X-ray and radio observations}.
\newblock {\em \aap}, 529:A66, May 2011.

\bibitem{Woo:1995aa}
R.~{Woo}, J.~W. {Armstrong}, M.~K. {Bird}, and M.~{Patzold}.
\newblock {Variation of fractional electron density fluctuations inside 40
  R$_{0}$ observed by ULYSSES ranging measurements}.
\newblock {\em \grl}, 22:329--332, February 1995.

\bibitem{Sasikumar-Raja:2017aa}
K.~{Sasikumar Raja}, P.~{Subramanian}, R.~{Ramesh}, A.~{Vourlidas}, and
  M.~{Ingale}.
\newblock {Turbulent Density Fluctuations and Proton Heating Rate in the Solar
  Wind from 9-20 Solar Radii}.
\newblock {\em \apj}, 850:129, December 2017.

\bibitem{Krupar:2018aa}
V.~{Krupar}, M.~{Maksimovic}, E.~P. {Kontar}, A.~{Zaslavsky}, O.~{Santolik},
  J.~{Soucek}, O.~{Kruparova}, J.~P. {Eastwood}, and A.~{Szabo}.
\newblock {Interplanetary Type III Bursts and Electron Density Fluctuations in
  the Solar Wind}.
\newblock {\em \apj}, 857(2):82, April 2018.

\bibitem{Bisoi:2014aa}
S.~K. {Bisoi}, P.~{Janardhan}, M.~{Ingale}, P.~{Subramanian},
  S.~{Ananthakrishnan}, M.~{Tokumaru}, and K.~{Fujiki}.
\newblock {A Study of Density Modulation Index in the Inner Heliospheric Solar
  Wind during Solar Cycle 23}.
\newblock {\em \apj}, 795:69, November 2014.

\end{thebibliography}

\section*{Data availability}
The simulation data sets generated during and/or analysed during the current study are available from the corresponding author on reasonable request.  LOFAR data used during the current study is publicly available on the LOFAR Long Term Archive at \url{https://lta.lofar.eu/} under the project codes LC3\_012 and LC4\_016.  The data used to make the plots in the paper is available on the UCL Research Data Repository using the DOI 10.5522/04/14140007

\section*{Code availability}
The code used to make the plots in the paper is available on the UCL Research Data Repository using the DOI 10.5522/04/14140679.  The code used to generate the data sets in our study is currently in preparation to be made publicly available.  In the interim period, the code can be made available from the corresponding author on reasonable request.

\end{document}